 \documentclass[preprint,10pt]{elsarticle} 

\usepackage[utf8]{inputenc}
\usepackage{graphicx}
\usepackage{amsmath}
\usepackage{subfig}
\usepackage{float}
\usepackage{cite}
\usepackage{amssymb}
\usepackage{mathtools}
\usepackage[symbol]{footmisc}

\usepackage[a4paper, total={6in, 8in}]{geometry}
\usepackage[toc,page]{appendix}

\usepackage[colorlinks,bookmarksopen,bookmarksnumbered,citecolor=blue,urlcolor=blue]{hyperref}
\usepackage[usenames,dvipsnames,svgnames,table]{xcolor}
\usepackage{tabularx}
\usepackage{booktabs}
\usepackage{array}
\newcolumntype{x}[1]{>{\centering\arraybackslash\hspace{-4em}}p{#1}}
\newcommand{\red}[1]{{\color{red} #1}}

 \usepackage{algorithm}
\usepackage{algpseudocode}

\renewcommand{\vec}{\mathbf}

\newcommand{\review}[1]{{\color{black} #1}}

 \usepackage{todonotes}
\begin{document}
\begin{frontmatter}
\title{
An Eulerian Meshless Method for Two-phase Flows with Embedded Geometries}


\author[inst1]{Anand S Bharadwaj\footnote{Corresponding author: anandbharadwaj1950@gmail.com}$^,$}
\author[inst2]{Pratik Suchde}
\author[inst1]{Prapanch Nair}

\affiliation[inst1]{organization={Indian Institute of Technology Delhi},
            addressline=	{Hauz Khas}, 
            postcode={110016}, 
            state={New Delhi},
            country={India}}

\affiliation[inst2]{organization={University of Luxembourg},
            addressline=	{2 avenue de l'universite}, 
            postcode={L-4365}, 
            state={Esch-sur-alzette},
            country={Luxembourg}}
\begin{abstract}
We present a novel Eulerian meshless method for two-phase flows with arbitrary embedded geometries. The spatial derivatives are computed using the meshless generalized finite difference method (GFDM). The sharp phase interface is tracked using a volume fraction function. 
The volume fraction is advected using a method based on the minimisation of a directional flux-based error. For stability, the advection terms are discretised using upwinding schemes. 
In the vicinity of the embedded geometries, the signed distance function is used to populate the surface of the geometries  to generate a body-conforming point cloud.  
Consequently, the points on the boundaries participate directly in the discretisation, unlike conventional immersed-boundary methods where they are either used to calculate momentum deficit (for example, continuous forcing) or conservation losses (for example, cut-cell methods). The boundary conditions are, therefore, directly imposed at these points on the embedded geometries, opening up the possibility for a discretisation that is body-conforming and spatially varying in resolution, while retaining the consistency of the scheme. We present benchmark test cases that validate the method for two-phase flows, flows with embedded boundaries and a combination of both. 
\end{abstract} 
\end{frontmatter}
\section{Introduction}
\subsection{Two-phase flows}
The simulation of two-phase flows in complex shaped domains, a ubiquitous phenomenon in industrial problems, aggregates different numerical challenges such as consistency, conservation, capturing of interface sharpness, satisfaction of no-slip at the immersed boundary  and spatial resolution. Often, the problems of complexly shaped solid boundaries and two-phase interface evolution are tackled separately; few solvers attempt to address both these phenomena in the same framework \cite{sun2016numerical,tavares2024coupled}. 
Traditionally, two-phase flows have been solved in Eulerian Cartesian mesh-based methods using a variety of techniques: front tracking, volume-of-fluid (VOF), level set (LS) and coupled level set-volume of fluid (CLSVOF) methods. While front-tracking methods \cite{unverdi1992front,popinet1999front,tryggvason2001front} capture interfacial phenomena sharply using Lagrangian markers, they are challenging for flows with topological changes such as break-up and coalescence.  In VOF, LS and CLSVOF, the phase interfaces are implicitly modelled using fields with their own transport equation. These methods can handle topological changes. 
In VOF \cite{hirt1981volume,rider1998reconstructing,roenby2016computational}, the advection of a phase volume fraction function, $\alpha \in [0,1]$, is solved using a finite volume discretization. The interface is reconstructed using simple line interface (SLIC)\cite{hirt1981volume}, piecewise linear (PLIC) \cite{rider1998reconstructing} or parabolic reconstruction (PROST) \cite{renardy2002prost}. The VOF method, despite showing good conservation properties, can be challenging in estimating interface curvature. The VOF methodology is also used to embed solid boundaries sharply within a fluid domain \cite{popinet2003gerris}. 
In the Level set method \cite{osher1988fronts,sussman1994level,sussman1998improved}, interfaces are tracked using level sets of a function (typically the signed distance function)  using the advection equation. While this method is more accurate for calculating curvature, for reasonable mass conservation, the \emph{level set function} needs to be reinitialized at regular intervals. To achieve conservation and accuracy, the coupled level set-volume of fluid (CLSVOF) method \cite{bourlioux1995coupled,sussman2000coupled,sussman2003second} was introduced. The level set function is reinitialized by using the volume fraction near the interface. A variation of the CLSVOF method was proposed by Tsui et al. \cite{tsui2013vof,tsui2017coupled} called the conservative interpolation scheme for interface tracking (CISIT), in which the interface is identified by the contour of the volume fraction field at $\alpha$ = 0.5. The level set function is then reinitialized with respect to this interface position. Since mesh generation and manipulation are expensive, the above approaches have been introduced in the context of Cartesian meshes. Adapting the resolution of Eulerian meshes and conforming the computational nodes to immersed bodies are two major challenges that limit their application.

Meshless methods are a widely used alternative to Eulerian mesh-based methods for simulating flows with interfaces. In Lagrangian meshless methods, the interfaces evolve naturally as the particles representing the phases move. The moving particle semi-implicit (MPS) \cite{koshizuka1996moving,koshizuka1998numerical,nomura2001numerical} method has been used to solve free-surface and two-phase flows. The method relies on moving particles that interact with other particles in the neighbourhood based on the particle number density. Smoothed particle hydrodynamics \cite{monaghan1994simulating,yang2020two,nair2014improved,nair2019simulations} is a popular method that has found widespread use for free-surface flows and two-phase flows. The generalized finite difference method \cite{veltmaat2022mesh,saucedo2019three,resendiz2018application,suchde2024meshfree}, in its Lagrangian form, is yet another meshless method that has found extensive use in two-phase flows and free-surface flows. In the Eulerian framework, the application of meshfree methods to free-surface and two-phase flows is relatively rare. The radial basis function 
(RBF) method has found application in an Eulerian framework to multiphase phase-change problems involving fluids. Abbaszadeh and Dehghan \cite{abbaszadeh2020simulation} apply the RBF method to the Shan-Chen model\cite{shan1993lattice}. Dehghan and 
 Najafi\cite{dehghan2016numerical} apply the RBF framework for liquid-solid phase change by solving the Stefan problem. In the work of Heydari et al. \cite{heydari2022new}  a Lagrangian-Eulerian approach using SPH has been proposed to solve a free-surface flow in an Eulerian framework away from the free-surface and a Lagrangian framework close to the free surface. However, the relevance of Eulerian meshless methods for problems with complex \emph{and} evolving interfaces is seldom explored. 

In the present work, a two-phase Eulerian meshless solver that captures interfaces in a manner similar to the above-mentioned CISIT method is introduced. The novelty of the method is that it leverages the advantages of the Eulerian as well as the meshless frameworks. While the Eulerian framework obviates the need to recalculate neighbourhoods of points at each time step, the meshless framework obviates the need for an apriori mesh. A volume fraction of one of the phases ($\alpha$) is declared as a field variable and the interface is located at $\alpha$=0.5. The volume fraction is advected with the flow and a directional flux-based error minimization scheme is proposed for the convective terms of the advection equation to be solved in the meshless framework. This scheme allows the use of upwinding in the directional fluxes used in the minimization procedure. Additionally, an interface sharpening method is proposed that makes use of the step nature of the volume fraction function and assists with the retention of the interface sharpness. For modelling surface tension, the level-set function needs to be reinitialized from an initial estimate that is constructed from the volume fraction such that zero-contour of the level-set function coincides with the interface i.e. the 0.5-contour of $\alpha$. 
\subsection{Methods for embedded geometries}
In mesh-based non-body conforming frameworks, immersed-boundary methods \cite{peskin1972flow} have been widely used to simulate flows past complex and moving geometries. They offer the advantage of not requiring a body-conformal mesh, thus, alleviating the difficulties of mesh generation. Consequently, to satisfy the no-slip and no-penetration boundary conditions at the embedded surface, forcing techniques are used. A variety of forcing techniques have been proposed depending on their suitability for specific flow problems. They can be broadly classified as - continuous and discrete forcing methods. In continuous forcing methods \cite{miller2009flexible}, a forcing function is added to the governing equations in the vicinity of the embedded surface such that the necessary boundary condition is satisfied. In discrete forcing methods \cite{tseng2003ghost}, cells in the vicinity of the embedded surface are flagged and the solution is reconstructed in these cells using a suitable interpolation such that the boundary condition at the embedded surface is satisfied. As with the case in forcing, to estimate the flow variables at the surface such as pressure, shear stress and the loads acting on the geometry, suitable interpolation strategies are necessary \cite{ghosh2020data}. 

In the present work, we present a method to accommodate objects in a flow as embedded surfaces in a non-conformal point cloud, in a manner similar to immerse-boundary methods. Conforming the point cloud to the embedded surface would be advantageous for an accurate representation of the shape of the surface as well as capturing the boundary layer effects.
To achieve this, we propose a method to generate a conformal point cloud on an arbitrary embedded geometry from a non-conforming initial arrangement of points. This can also be applied to moving geometries. Thus the surface participates in the computation of spatial gradients without the need for interpolation. 
This ensures direct enforcement of boundary conditions, in contrast to the introduction of source terms in the momentum equation in immersed-boundary methods. Additionally, the surface quantities such as pressure and shear stresses can be extracted directly from the surface points, without a need for  interpolation.  

 The paper is organized as follows. Sec.~\ref{sec:methodology} discusses methodology -- the governing equations, the Generalized Finite Difference Method (GFDM), the interface tracking algorithm and the generation of conformal point clouds for arbitrary embedded surfaces. Sec.~\ref{sec:results} presents different test cases for validating the model. Finally, Sec.~\ref{sec:conclusions} presents the conclusions and indicates some future extensions of the proposed model. 
 \section{Methodology}\label{sec:methodology}
In this section, we present the new Eulerian meshless two-phase flow solver with embedded geometries. We present the governing equations, followed by the generalized finite difference method (GFDM) that discretises the governing equations. Subsequently, we propose the interface tracking algorithm with emphasis on an interface sharpening method and directional flux-based minimization in the solution to the volume fraction advection equation. Finally, we propose a method that conforms a point cloud in a regular lattice onto an arbitrary geometry. 
\subsection{Governing equations}
The incompressible Navier-Stokes equations are considered : 
\begin{equation}
\nabla \cdot \mathbf{u} = 0 \text{ , }
    \label{eq:cont}
\end{equation}
\begin{equation}
    \frac{\partial\mathbf{u}}{\partial t} + (\mathbf{u} \cdot \nabla) \mathbf{u} = -\frac{1}{\rho}\nabla p + \nu \Delta \mathbf{u} + \mathbf{g} \text{ , }
    \label{eq:mom}
\end{equation}

where $\mathbf{u}$ denotes the velocity, $p$, the pressure, $\mathbf{g}$, the acceleration due to gravity, $\rho$, the density and $\nu$, the kinematic viscosity, respectively. \\ 
The projection method \cite{chorin1968numerical} is used to solve the above equations.
The momentum equations are first marched to solve for a provisional velocity field. 
\begin{equation}
    \frac{\mathbf{u}_i^* - \mathbf{u}_i^n}{\Delta t}  = - \mathbf{F}_c|^n  + \mathbf{F}_v|^n + \mathbf{g}.
    \label{eq:mom_discrete}
\end{equation}

The convective and viscous terms are denoted as $\mathbf{F}_c$ and $\mathbf{F}_v$ respectively. The superscript `$n$' denotes the time level and `*' denotes the provisional values. 
The provisional velocity field generally does not satisfy the divergence-free condition. The pressure $p$ is coupled to the velocity through the Poisson equation:
\begin{equation}
    \nabla^2 p^{n+1} = \frac{\rho}{\Delta t} \nabla \cdot 
    \mathbf{u}^* + \frac{1}{\rho} \nabla p^{n+1} \cdot \nabla \rho .
    \label{eq:pressurepoisson}
\end{equation}
The second term on the RHS of the above equations may be non-zero in situations where two phases of different densities exist in the flow giving rise to a non-zero density gradient in the vicinity of the interface. Having solved the above pressure Poisson equation, the velocity is then corrected as
\begin{eqnarray}
    \mathbf{u}^{n+1} &=& \mathbf{u}^* - \frac{\Delta t}{\rho} \nabla p^{n+1} \text{ .}
\end{eqnarray}
\subsection{Generalized Finite Difference Method}
\begin{figure}[htb]
    \centering
    \includegraphics[width=0.5\textwidth]{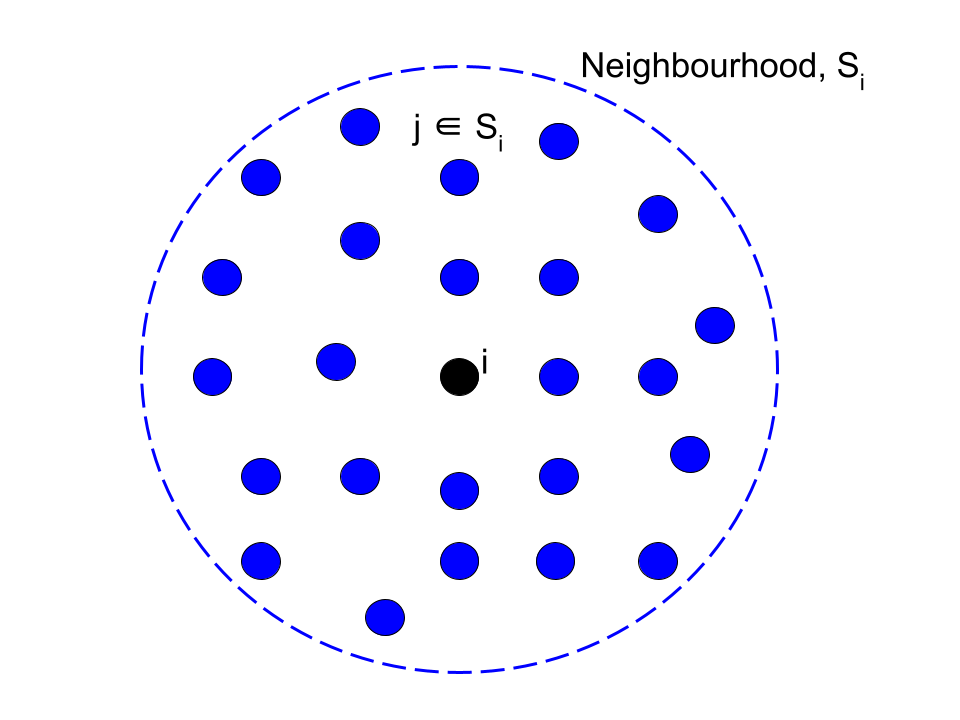}
    \caption{Generalized Finite Difference Method: Neighbourhood of a point in the point cloud.}
    \label{fig:GFDM}
\end{figure}

The generalized finite difference method (GFDM) is a meshless method that estimates derivatives of flow variables at a point from a set of neighbours that are a part of the point cloud using a weighted least squares error minimization procedure as discussed below \cite{seibold2006m,suchde2018conservation,benito2001influence,benito2007solving,gavete2003improvements,rao2023upwind,clain2024stencil}. \\
Consider a point $i$ which has a neighbourhood of points, $j \in S_i$. $S_i$ denotes the support region around the point $i$. This is illustrated in Fig.~\ref{fig:GFDM}.
Using a monomial basis up to a prescribed degree in two spatial dimensions, such as 

\begin{equation}
M_i(x,y) =
   \begin{bmatrix}
1 & \Delta x & \Delta y & \Delta x^2 & \Delta y^2 & \Delta x \Delta y \\
\end{bmatrix} \text{, }
\label{eq:monomials}
\end{equation}
 differential operators can be derived for a non-uniformly discretized field. 
Here, $\Delta x = x - x_i$ and $\Delta y = y - y_i$. 
As an example, we derive the procedure for the Laplacian operator.  Let us denote the Laplacian operator at a point $j$ in the neighbourhood of point $i$  as $C^\Delta_{ij}$. Applying the Laplacian operator to each of the monomial basis elements in Eq.~\ref{eq:monomials}, we get
\begin{equation}
	\begin{aligned}
		&\sum_{j \in S_i} C^\Delta_{ij} (1) &= 0& \text{, } &   \sum_{j \in S_i} C^\Delta_{ij} (\Delta x_j) &= 0 \text{, }\\
		&\sum_{j \in S_i} C^\Delta_{ij} (\Delta y_j) &= 0& \text{, } &    \sum_{j \in S_i} C^\Delta_{ij} (\Delta x_j^2) &= 2 \text{, }\\
		&\sum_{j \in S_i} C^\Delta_{ij} (\Delta y_j^2) &= 2&\qquad\text{ and } &    \sum_{j \in S_i} C^\Delta_{ij} (\Delta x_j \Delta y_j) &= 0, \\
\end{aligned}
\label{eq:consistency}
\end{equation}
which become the consistency conditions for the operator. This can be rewritten concisely as, 
\begin{equation}
    \mathbf{V}_i \vec{C^\Delta}_i = \vec{b}_i\text{, }
    \label{eq:constraints}
\end{equation}
where,
\begin{equation}
    \mathbf{V}_i =
    \begin{pmatrix}
        \ldots & 1          &  \ldots \\
        \ldots & \Delta x_j &   \ldots \\
        \ldots & \Delta y_j &    \ldots\\
        \ldots & \Delta x_j^2 &   \ldots\\
        \ldots & \Delta y_j^2 &    \ldots \\
        \ldots & \Delta x_j \Delta y_j&   \ldots \\
    \end{pmatrix}
, \, 
\vec{C^\Delta}_i = 
    \begin{pmatrix}
        C^\Delta_{i1} \\
        \vdots \\
        C^\Delta_{ij} \\
        \vdots \\
        C^\Delta_{iN} \\ 
    \end{pmatrix} 
    \, \text{and }
 \vec{b}_i =
    \begin{pmatrix}
        0 \\
        0 \\
        0 \\
        2 \\
        2 \\
        0 \\
    \end{pmatrix}.
\end{equation}
Henceforth we drop the index $i$ for the tensorial quantities, for brevity, as the ensuing discussion concerns a given particle $i$. 
Eq.~\ref{eq:constraints} is
used in minimizing the functional
\begin{equation}
    J = \sum_{j \in S_i} \frac{(C^\Delta_{ij})^2}{w_{ij}}.
\end{equation}
Following the weighted least squares procedure, we get
\begin{equation}
    \vec{C^\Delta} = \mathbf{W} \mathbf{V}^T \left(\mathbf{V} \mathbf{W} \mathbf{V}^T\right)^{-1} \vec{b}
    \label{eq:op_formula}
\end{equation}
where $\mathbf{W}$ is a diagonal matrix with its entries given by the weight of the point $j \in S_i$ w.r.t. the point $i$, 
\begin{equation}
	\mathbf{W} = 
\begin{pmatrix}
	w_{i1}& \ldots& 0 \\
	\vdots&\ddots&\vdots \\
	0 &\ldots & w_{iN} 
\end{pmatrix}.
\end{equation}  
In this work, the weights are assigned using the Gaussian function. The weight of a point $j$ in the neighbourhood of a point $i$ is 
\begin{equation}
    w_{ij} = \frac{1}{\pi h^2} \exp\left(-\frac{|\vec{r}_i  - \vec{r}_j|^2}{h^2}\right)\text{, }
\end{equation}
where $\vec{r}_i$ and $\vec{r}_j$ are the position vectors of point $i$ and $j$, respectively, and $h$ is the smoothing length which is typically chosen to be the radius of the circle in Fig.~\ref{fig:GFDM}. \review{The choice of the weight function does not significantly alter the numerical results according to \cite{jacquemin2020taylor}, and this is observed in our simulations as well.}\\
Having derived the Laplacian operator from Eq.~\ref{eq:op_formula}, the Laplacian of a general function $\phi$ at the point $i$  may be approximated as
\begin{equation}
    \nabla^2 \phi |_i \approx \sum_{j \in S_i} C^\Delta_{ij} \phi_j\text{. }
    \label{eq:laplacianphi}
\end{equation}
The process is identical the for other operators. For the first order derivatives in $x$ and $y$, the operator $C^\Delta_{ij}$ in Eq.~\ref{eq:laplacianphi} is simply replaced with the corresponding operators--- $C^x_{ij}$ and $C^y_{ij}$, respectively. \review{It is noted that the operators can also be derived by minimizing the truncation error of the Taylor series. The approach using Taylor series and the approach using monomials (as detailed above) are mathematically equivalent \cite{suchde2018conservation}.} \review{ For work on higher order operators in GFDM, readers are referred to \cite{clain2024stencil,qu2024stable,sun2024hybrid,milewski2012meshless,kraus2023higher}.} \\
We, now, look at the discretization of the convective terms ($\mathbf{F}_c$) from the Navier-Stokes equations (Eqs.~\ref{eq:mom_discrete}).
From the $x$-momentum equation, the convective terms are 
\begin{equation}
    F_c^x = \frac{\partial f}{\partial x} + \frac{\partial g}{\partial y}\text{, }
\end{equation}
where $f = u^2$ and $g = uv$.  
The derivatives are computed as
\begin{equation}
    \frac{\partial f}{\partial x} = \sum_{j \in S_i} C^x_{ij} f_j, \, \frac{\partial g}{\partial y} = \sum_{j \in S_i} C^y_{ij} g_j  .
\end{equation}
The same procedure is repeated for the convective terms of the $y$-momentum equation. \\
The viscous term $\mathbf{F}_v$ involves the Laplacian of the velocity components. 
\begin{equation}
    F_v^x = \nu \nabla^2 u = \nu \sum_{j \in S_i} C^\Delta _{ij} u_j \text{, }
\end{equation}
\begin{equation}
    F_v^y = \nu \nabla^2 v = \nu \sum_{j \in S_i} C^\Delta _{ij} v_j \text{. }
\end{equation}
For the pressure Poisson equation, the Laplacian operator is applied to the pressure field in the same manner as above. In the discretized form, the pressure Poisson equation (Eq.~\ref{eq:pressurepoisson}) can be written as
\begin{equation}
\begin{aligned}
    \sum_{j \in S_i} C^\Delta_{ij} p^{n+1}_j = & \frac{\rho_i}{\Delta t} \sum_{j \in S_i} \left[C^x_{ij} u^ *_j +  C^y_{ij} v^*_j \right]+ \\
    & \frac{1}{\rho_i} \left[\sum_{j \in S_i} C^x_{ij} p^{n+1}_j \sum_{j \in S_i} C^x_{ij} \rho_j + \sum_{j \in S_i} C^y_{ij} p^{n+1}_j \sum_{j \in S_i} C^y_{ij} \rho_j  \right]\text{. } \\
\end{aligned}
\end{equation}
At the boundaries, Dirichlet boundary conditions are imposed by simply setting the pressure or velocity to the prescribed value. On the other hand, Neumann boundary conditions are imposed using the differential operators. As an example, let's impose a zero-Neumann condition for pressure at a boundary point $b$ with a boundary normal $\vec{n}^b = (n_x^b, n_y^b)$ such that $\nabla p_b \cdot \vec{n}^b = 0$.
\begin{equation}
    \nabla p_b \cdot \vec{n}^b = n_x^b \sum_{j \in S_b} C^x_{bj} p_j + n_y^b \sum_{j \in S_b} C^y_{bj} p_j = 0\text{. } 
\end{equation}
We rearrange the terms to get an expression for pressure at the boundary as
\begin{equation}
    p_b = -\frac{ n_x^b \sum_{j \in S_b \atop j \neq b} C^x_{bj} p_j + n_y^b \sum_{j \in S_b \atop j \neq b} C^y_{bj} p_j} { n_x^b C^x_{bb} +  n_y^b C^y_{bb} }\text{. }
\end{equation}
\subsection{Interface tracking for two-phase flows}
The interface tracking method used in this work is inspired from the CLSVOF-bsaed CISIT technique \cite{tsui2013vof,tsui2017coupled}. At each point of the point cloud, a volume fraction is defined as below: 
\begin{equation}
    \alpha = \left \{
    \begin{array}{c}
        1 \text{, if $i$ is in phase 1}   \\
        0 \text{, if $i$ is in phase 2}\\ 
    \end{array} \right. .
\end{equation}
At the vicinity of the interface between the two phases, $\alpha$ varies smoothly from $\alpha=0$ to $\alpha=1$ and the interface location is identified as $\alpha = 0.5$. The volume fraction is advected at the velocity of the flow 
\begin{equation}
    \frac{\partial \alpha}{\partial t} +   \nabla \cdot (\mathbf{u}  \alpha) = 0 .
    \label{eq:alpha_advect}
\end{equation}
The flux is determined by a directional flux-based minimization procedure described in Sec.~\ref{sec:fluxminimization}. In case surface tension terms are included in the momentum equation, a level set (signed distance) function, $\phi$, is required to estimate the surface curvature. In the case of flows with surface tension (not dealt with in this paper), the level set function needs reinitialization (as provided in \cite{sussman1998improved}, for example) for mass conservation. 
\subsubsection{Interface sharpening method}
\label{sec:sharp}
While advecting the volume fraction as per Eq.~\ref{eq:alpha_advect}, a hyperbolic partial differential equation, the discretization has a tendency to diffuse the sharp change of $\alpha$ at the interface owing to dissipative errors in upwind schemes. Higher order discretizations such as TVD with limiting and WENO schemes preserve the sharpness better than first-order schemes, although a certain amount of dissipation is inevitable. Here, we propose a method to preserve the sharpness at the interface, in addition to the choice of the discretization. \\
The method relies on the fact that the volume fraction, $\alpha$, is  a step function, in principle, and therefore, the possible values it can assume are either 0 or 1, with a narrow region at the interface where it changes smoothly from 0 to 1. This narrow region widens due to the effect of dissipative numerical errors. Let us define a step function, $\psi'$, such that 
\begin{equation}
    \psi' = 1- 2 \alpha \text{. }
\end{equation}
$\psi'$, therefore, varies linearly from -1 to 1 in the vicinity of the interface, with its extent of diffusion same as that for $\alpha$. Further, we define $\psi$ such that
\begin{equation}
    \psi = \text{sign}(\psi')\text{. }
\end{equation}
$\psi$ takes on the values -1 and 1 and is discontinuous at the interface. \\
We now use the smoothing operation as a filter on $\psi$ at each point $i$, as defined below, such that the sharp discontinuity gets diffused over a narrow region over which it smoothly varies from -1 to 1.
\begin{equation}
\displaystyle
    \psi_i^{(1)} = \frac{\sum_{j \in S_i} w_{ij} \psi_j}{\sum_{j \in S_i} w_{ij}}\text{, }
\end{equation}
where $w_{ij}$ is a weight used for the smoothing. This smoothing is necessary to maintain numerical stability. Also, the smoothed region is narrower than the diffused region that forms due to the dissipative errors, thus making it a sharper representation of the interface.
Further, the volume fraction is revised as
\begin{equation}
    \alpha = \frac{(1-\psi^{(1)})}{2}\text{. }
\end{equation}
This method for interface sharpening can be done at regular intervals as the flow evolves. However, doing it too often can lead to an erroneous evolution of the solution. In the results section, we assess the effect of frequency of interface sharpening on the solution for the Rayleigh-Taylor instability. 
\subsubsection{Directional flux-based minimization}\label{sec:fluxminimization}
\begin{figure}[htb]
    \centering
    \includegraphics[width=0.5\textwidth]{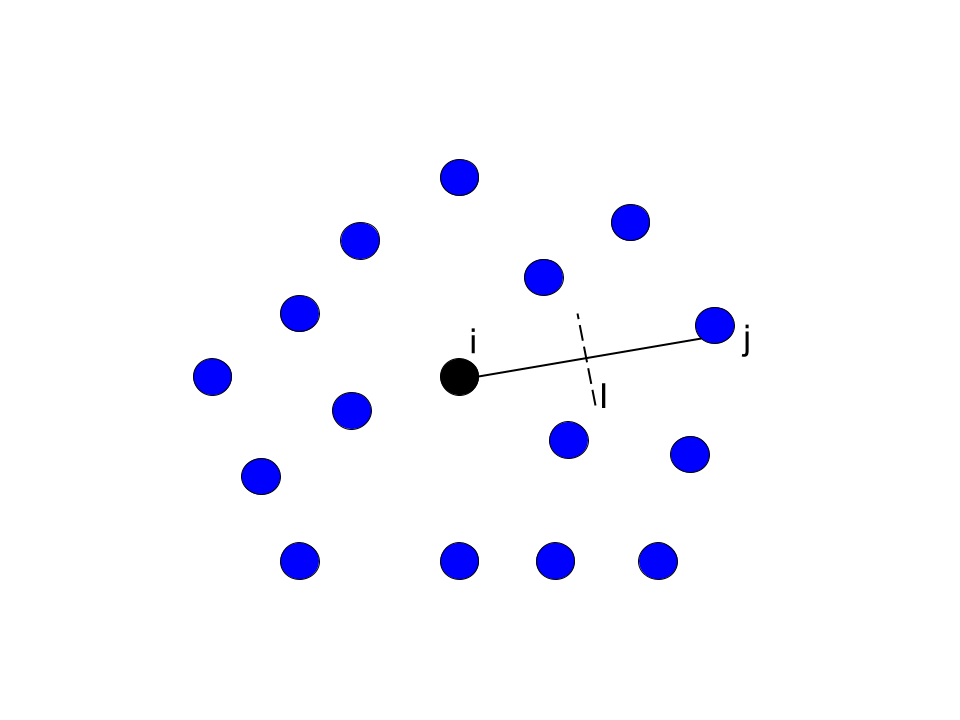}
    \caption{Fictitious interface (I) between a point $i$ and any neighbour, $j$.}
    \label{fig:GFDM_interface}
\end{figure}

Let us consider the convective term of Eq.~\ref{eq:alpha_advect} 
\begin{equation}
    \nabla \cdot (\mathbf{u} \alpha) = \nabla \cdot \mathbf{F} \text{. }
\end{equation}
The flux vector 
$$\mathbf{F} = \begin{bmatrix} f &   g \end{bmatrix}^T \text{, }$$ 
where $f = u\alpha$ and $g = v\alpha$. In contrast to finite-volume methods, the meshless method presented here does not deal with cells and their faces as a part of the discretization. Thus, to estimate a flux, we consider a fictitious interface between a point $i$ and its neighbour point $j$, as shown in Fig.~\ref{fig:GFDM_interface}.\\
The directional flux at the interface $I$ (for each $i$--$j$ pair) with unit normal $\hat{\mathbf{n}}$ is given by 
\begin{equation}
    \mathbf{F}_I \cdot \hat{\mathbf{n}} = \begin{bmatrix} f_I & g_I \end{bmatrix}^T \cdot \hat{\mathbf{n}} 
    \label{eq:fdotn}
\end{equation}
The unit normal $\hat{\mathbf{n}}$ is the unit vector pointing from point $i$ to point $j$.
Expanding the terms $f_I$ and $g_I$ w.r.t point $i$ using Taylor Series, we get
\begin{equation}
    f_I = f_i + f_{x,i} \Delta x + f_{y,i} \Delta y + e_f\text{, }
\end{equation}
\begin{equation}
    g_I = g_i + g_{x,i} \Delta x + g_{y,i} \Delta y + e_g\text{, }
\end{equation}
where $f_{x,i}$ and $f_{y,i}$ are the partial derivatives of $f$ w.r.t $x$ and $y$ at the point $i$, respectively, $g_{x,i}$ and $g_{y,i}$ are the partial derivatives of $g$ w.r.t $x$ and $y$ at the point $i$, respectively, and $e_f$ and $e_g$ are the higher order terms in the series expansion. Substituting these expressions in Eq.~\ref{eq:fdotn}, we get
\begin{equation}
    F_I \cdot \hat{\mathbf{n}} = (f_i n_x + f_{x,i} \Delta x n_x + f_{y,i} \Delta y n_x) + (g_i n_y + g_{x,i} \Delta x n_y + g_{y,i} \Delta y n_y ) - e
\end{equation}
Here, $e$ is the accumulated higher order term. Rearranging the above equation, 
\begin{equation}
    e = f_i n_x + g_i n_y + f_{x,i} \Delta x n_x + g_{y,i} \Delta y n_y  - (F_I \cdot \hat{\mathbf{n}} - f_{y,i} \Delta y n_x - g_{x,i} \Delta x n_y )
    \label{eq:error}
\end{equation}
For a given point $i$ (dropping the subscript henceforth, for tensorial terms), there will be $N$ such equations corresponding to $N$ neighbours such that 
\begin{equation}
    \mathbf{E} =     \begin{bmatrix} e_1 &\cdots & e_j &\cdots & e_N \end{bmatrix}^T
         = \mathbf{M}\mathbf{a} - \mathbf{d}\text{, }
\end{equation}
where 
\begin{equation}
     \mathbf{M} =     
    \begin{bmatrix}
        n_{x,1} & n_{y,1} & \Delta x_1 n_{x,1} & \Delta y_1 n_{y,1} \\
         &  & \vdots &  \\
         n_{x,j} & n_{y,j} & \Delta x_j n_{x,j} & \Delta y_j n_{y,j} \\    
         &  & \vdots &  \\
         n_{x,N} & n_{y,N} & \Delta x_N n_{x,N} & \Delta y_N n_{y,N}
    \end{bmatrix}
    ,\, 
    \mathbf{a} = 
    \begin{bmatrix}
        f_i \\
        g_i \\ 
        f_{x,i} \\ 
        g_{y,i} \\
    \end{bmatrix}
    \text{ , }
    \mathbf{d} = 
        \begin{bmatrix}
        d_1 \\ 
        \vdots \\
        d_j \\
        \vdots \\
        d_N \\
    \end{bmatrix}.
\end{equation}

Here, $\mathbf{d}$ is a $N\times 1$ tensor, whose elements are $d_j = (F_I \cdot \hat{\mathbf{n}} - f_{y,i} \Delta y n_x - g_{x,i} \Delta x n_y )$, where all the quantities are evaluated between the point $i$ and neighbour $j$.  Standard interface solution reconstruction schemes developed for the finite-volume framework can be directly applied to the term $F_I \cdot \hat{\mathbf{n}}$. For the terms $f_{y,i}$ and $g_{x,i}$, the regular differential operators are used as shown below.
\begin{eqnarray}
    f_{y,i} = \sum_{j \in S_i} C^y_{ij} f_j \\\text{, }
    g_{x,i} = \sum_{j \in S_i} C^x_{ij} g_j \text{. }
\end{eqnarray}
We, now, minimize the below functional ($J$) w.r.t $\mathbf{a}$ for a given point $i$:
\begin{equation}
J = \mathbf{E}^T \mathbf{E}.
\end{equation}
The minimization leads to  
\begin{equation}
    \mathbf{a} = (\mathbf{M}^T \mathbf{M})^{-1}\mathbf{ M}^T  \mathbf{d}. 
\end{equation}
We use the third ($f_{x,i}$) and the fourth ($g_{y,i}$) elements of the solution vector $\mathbf{a}$ in the solution to the advection equation Eq.~\ref{eq:alpha_advect}.
\subsection{Generation of point clouds conforming to embedded geometries} \label{sec:IB}
\begin{figure}[htb]
    \centering
    \subfloat[]{
    \includegraphics[width=0.5\textwidth]{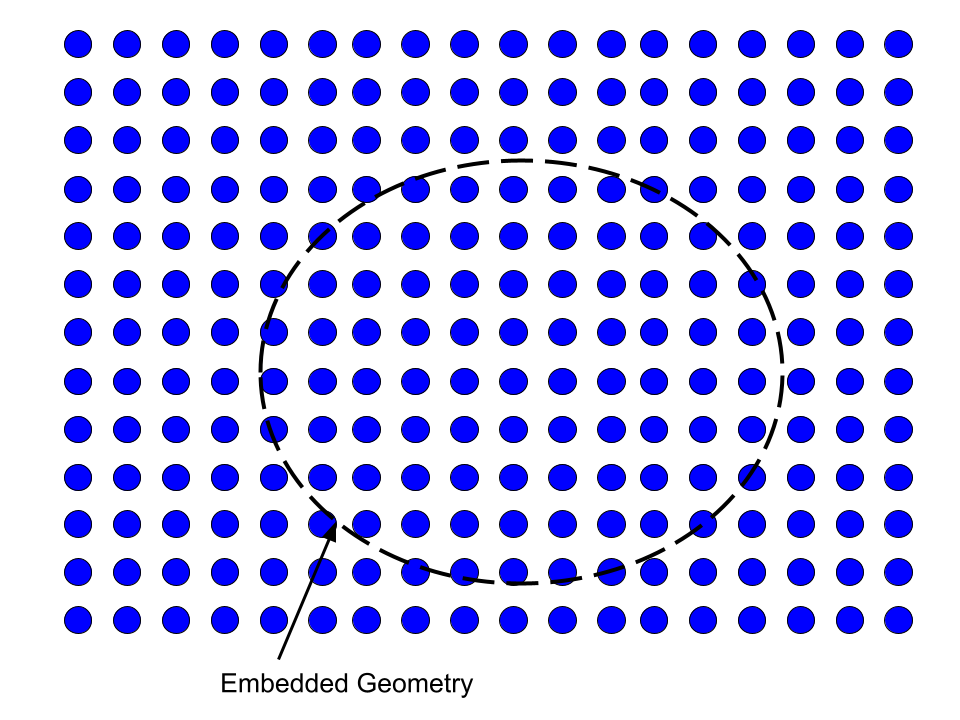}}
        \subfloat[]{
    \includegraphics[width=0.5\textwidth]{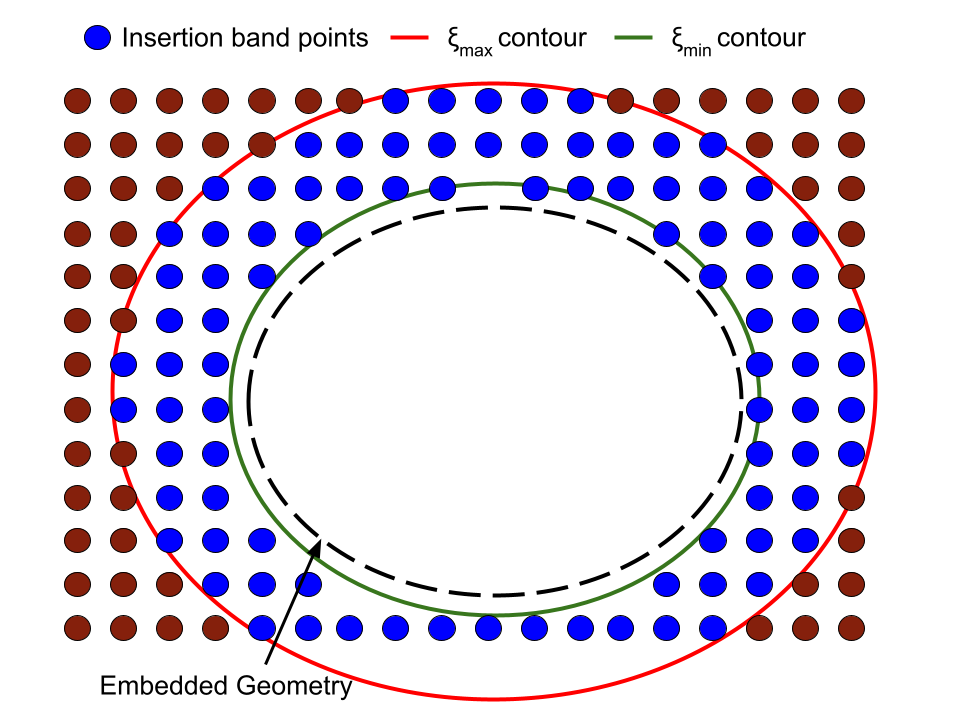}}
    \caption{Generation of conformal point clouds (a) A set of uniformly distributed points with an embedded geometry (b) $\phi_{\text{max}}$ and $\phi_{\text{min}}$ contours based on which the insertion band points are flagged. }
    \label{fig:IB}
\end{figure}

The present method starts with a non-conformal point cloud much like Cartesian meshes used in mesh-based immersed-boundary methods. However, in contrast to immersed-boundary methods, we introduce additional points on the surface of the embedded geometry which participate in the discretization of the governing equations directly. This is in contrast to the continuous forcing used in the immersed-boundary methods, where fields are interpolated at the immersed points. This provides a way to impose the boundary conditions exactly at the embedded surface. The points within the geometry are discarded for the case of stationary embedded geometry, reducing the memory footprint of the solver. \\
The procedure to form a conformal point cloud is elaborated here. We begin with a set of points that do not conform to the geometry, as shown in Fig.~\ref{fig:IB}(a).
The geometry is identified by a set of marker points (that are of relatively higher resolution than the point cloud) and the surface normal at each of these marker points. A signed distance function is defined for the points in the point cloud
\begin{equation}
    \xi_{i,EG} = |\vec{X}_i - \vec{X}_m| \text{ sign}((\vec{X}_i - \vec{X}_m)\cdot \vec{n}_m) \text{, }
    \label{eq:IB_sdf}
\end{equation}
where $\vec{X}_i$ and $\vec{X}_m$ are the positions of the point $i$ of the point cloud and the marker point $m$, respectively, and $\vec{n}_m$ is the unit outward surface normal at the marker  point $m$.\\
To populate points on the surface of the embedded geometry, a band of points called the insertion band, belonging to the point cloud, is considered such that the signed distance function in the insertion band lies in the range 
\begin{equation}
    \xi_{\text{min}} < \xi_{i,EG} < \xi_{\text{max}}\text{. }
\end{equation}
Here, $\xi_{\text{min}}$ is set to the order of the point cloud resolution near the embedded surface and $\xi_{\text{max}}$ is set to a value that is larger than $\xi_{\text{min}}$ by a factor ($\sim 5$, based on trial and error). In Fig.~\ref{fig:IB}(b), the insertion band is represented by blue points that lie within the contour lines of $\xi_{\text{min}}$ and $\xi_{\text{max}}$.\\
For each point $i$ in the insertion band, a corresponding surface point $s$ is constructed as
\begin{equation}
    \vec{X}_s = \vec{X}_i - \xi_{i,EG} \frac{\nabla \xi_{i,EG}}{|\nabla \xi_{i,EG}|}\text{. }
\end{equation}
Here, $\nabla \xi_{i,EG}/|\nabla \xi_{i,EG}|$ is the normal to the embedded geometry that passes through the point $i$.\\
This surface point, $s$, is appended to the point cloud if it lies at a distance of at least $\xi_{\text{min}}$ from all the points in the insertion band as well as the previously appended surface points. This is essential to avoid blowing up of the differential operators when the two points of the point cloud are too close.\\
The procedure outlined here provides a general framework for moving and deforming embedded geometries since the creation of the surface points relies on the signed distance function, $\xi_{EG}$, which evolves according to
\begin{equation}
    \frac{\partial \xi_{EG}}{\partial t} + \mathbf{u} \cdot \nabla \xi_{EG} = 0.
\end{equation}
In scenarios where only stationary bodies and rigid-body motion are  involved, the Lagrangian markers can directly be used to populate the surface of the embedded geometries, without the need for the use of the insertion band. 
All points $k$ of the non-conformal point cloud that satisfy $$\xi_{k,EG} < \xi_{\text{min}}$$ are discarded in the case of stationary geometry or temporarily deactivated in case of moving/deforming geometries. Even though the test cases presented here involve only stationary rigid bodies, we use the insertion band approach for the sake of generalisation. \\

\section{Results}\label{sec:results}
In this section, we present test cases to demonstrate different features of the solver, as remarked in Table.~\ref{tab:test_cases}. 
\begin{table}[htb]
    \centering
    \caption{Test cases}
    \begin{tabularx}{\textwidth}{clX}
    \toprule 
    Case & Description & Remarks \\
    \midrule
       1  &  Heat equation on an irregular domain & Quantification of error\\
       2  & Flow past a circular cylinder & Tests algorithm for embedded geometries \\
       3 & Rayleigh-Taylor instability & Tests the two-phase flow model\\
       4 & Two-phase Dam break & Tests two-phase flow model for high density ratio \\
       5 & Filling of a mould with core & Tests the two-phase flow model with embedded geometries \\
    \bottomrule
    \end{tabularx}
    \label{tab:test_cases}
\end{table}

\begin{figure}[htb]
    \centering
    \includegraphics[width=3.5in]{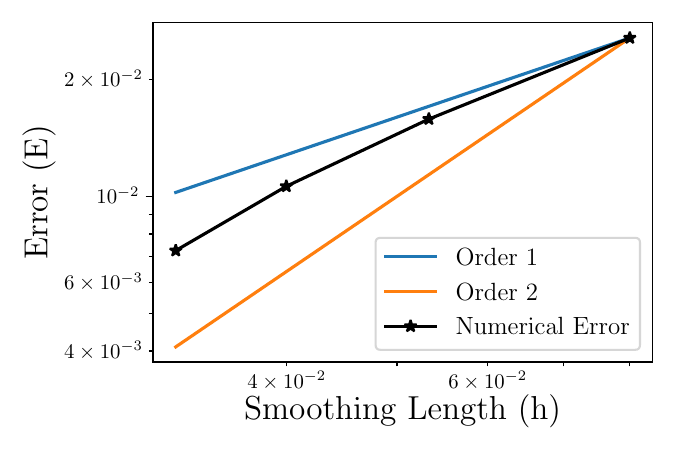}
    \caption{Case 1: Error convergence plot. \review{The smoothing lengths used are 0.032, 0.04, 0.053, 0.08 (from left to right on the plot)}}
    \label{fig:Err_convergence}
\end{figure}
\begin{figure}[htb]
    \centering
    \subfloat[Coarse point cloud, $PC_2$ ]{
	    \includegraphics[width=0.33\textwidth,clip,trim={7cm 0 4cm 0}]{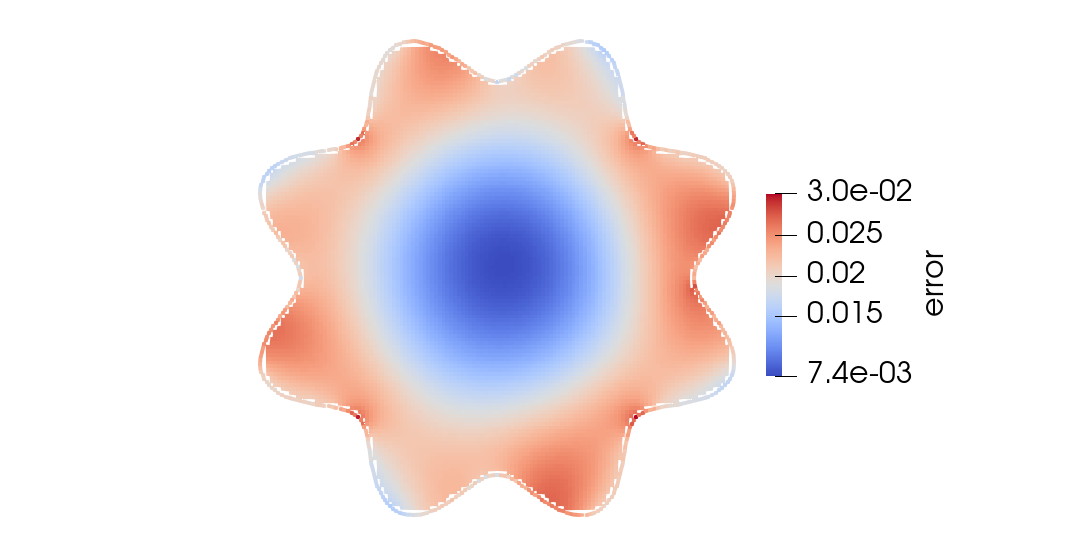}
    }
    \subfloat[Medium point cloud, $PC_3$]{
    \includegraphics[width=0.33\textwidth,clip,trim={7cm 0 5cm 0}]{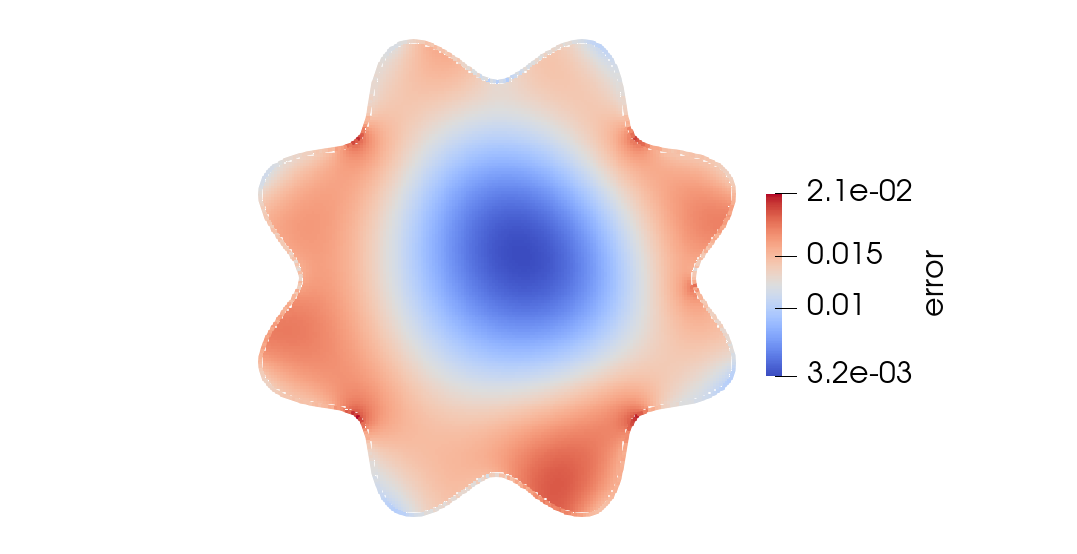}
    }
    \subfloat[Fine point cloud, $PC_4$]{
    \includegraphics[width=0.33\textwidth,clip,trim={7cm 0 5cm 0}]{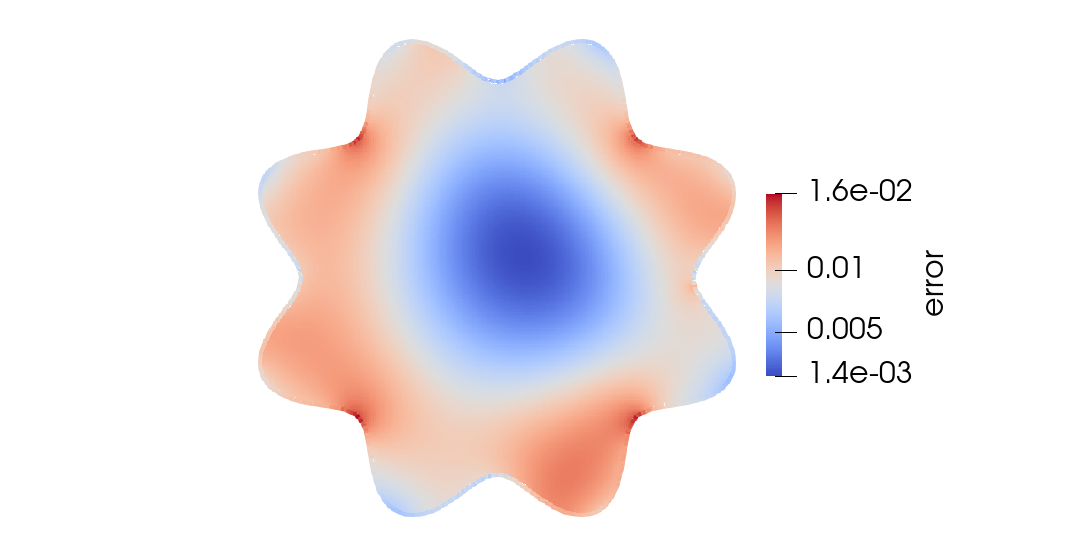}
    }
    \caption{Case 1: Error distribution in the irregular domain. The color bars show the error.}
    \label{fig:C1_error}
\end{figure}
\begin{figure}[htb]
    \centering
    \includegraphics[width=0.5\textwidth]{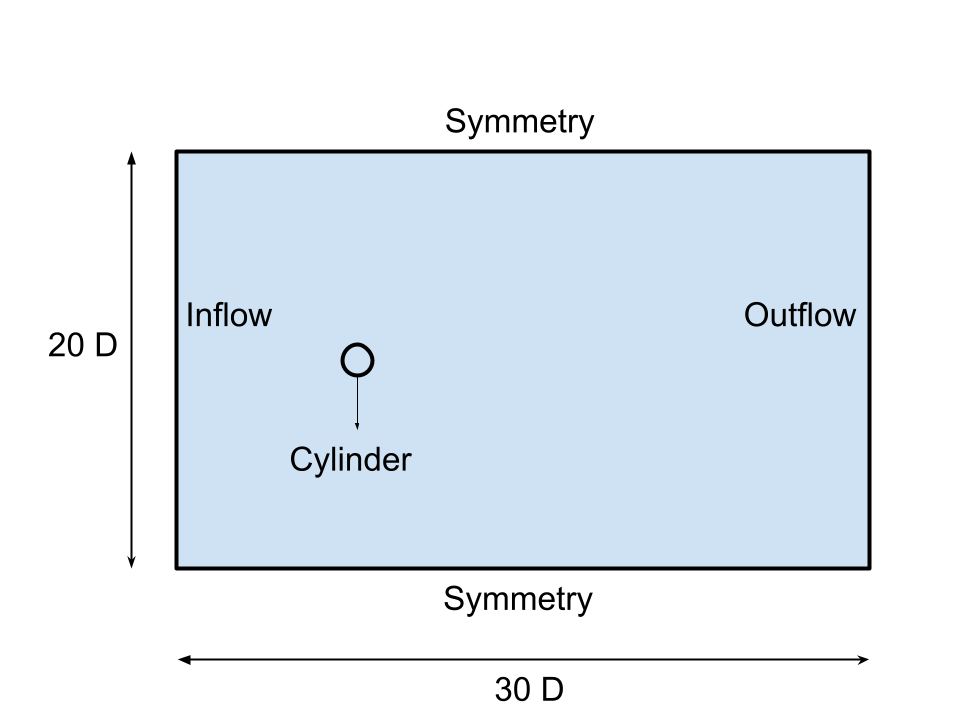}
    \caption{Case 2: Domain and boundary conditions for flow past a circular cylinder.}
    \label{fig:c2_domain}
\end{figure}
\begin{figure}[htb]
    \centering
    \subfloat[]{
	    \includegraphics[width=0.5\textwidth,clip,trim={6cm 1.5cm 6cm 0}]{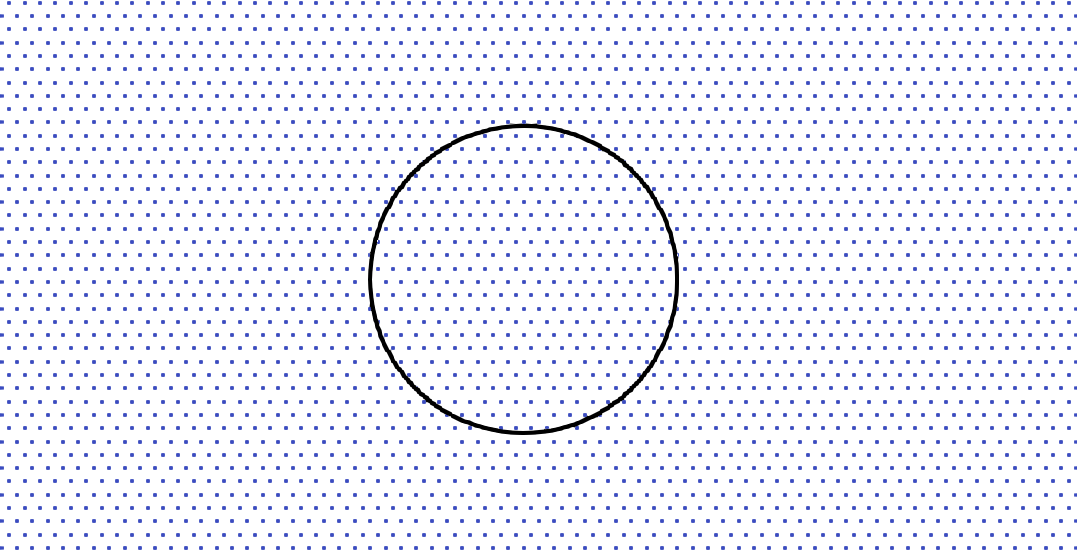}
    }\\
    
    \subfloat[]{
	    \includegraphics[width=0.5\textwidth,clip,trim={6cm 1.5cm 6cm 0}]{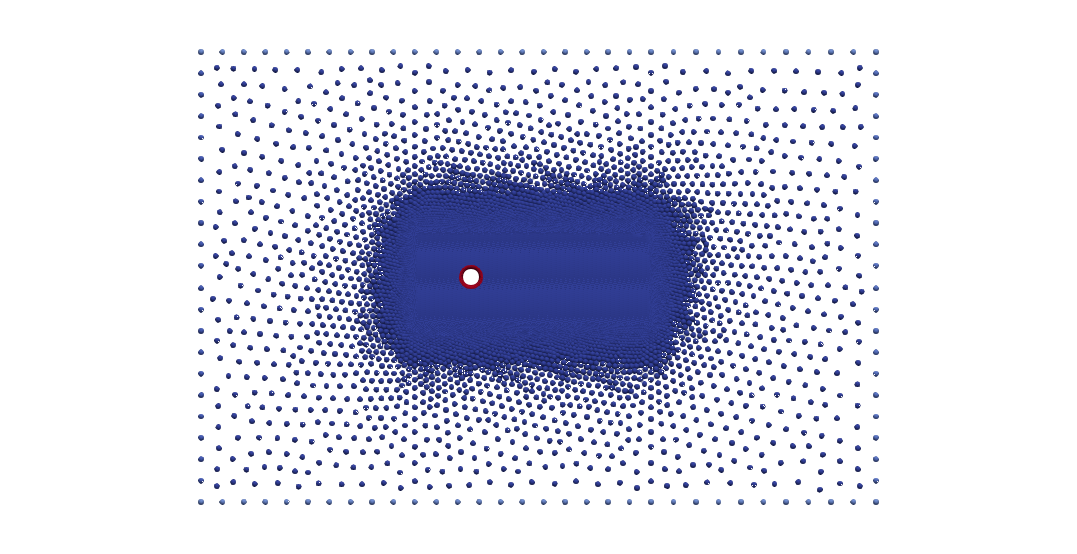}
    }
    \subfloat[]{
    \includegraphics[width=0.5\textwidth,clip,trim={6cm 1.5cm 6cm 1.5cm}]{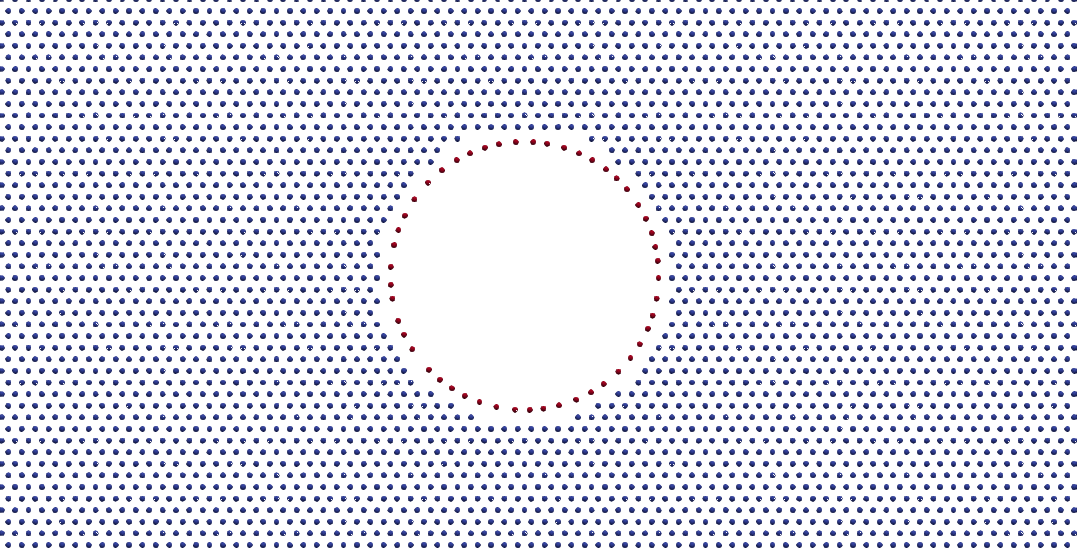}
    }
    \caption{Case 2: Flow past a cylinder: (a) The non-conformal point cloud with the cylinder as an embedded surface (b) The entire conformal point cloud (c) Zoomed-in view of the conformal point cloud near the cylinder }
    \label{fig:C2_pointcloud}
\end{figure}

\begin{figure}[htb!]
    \centering
    \subfloat[]{
    \includegraphics[width=0.55\textwidth]{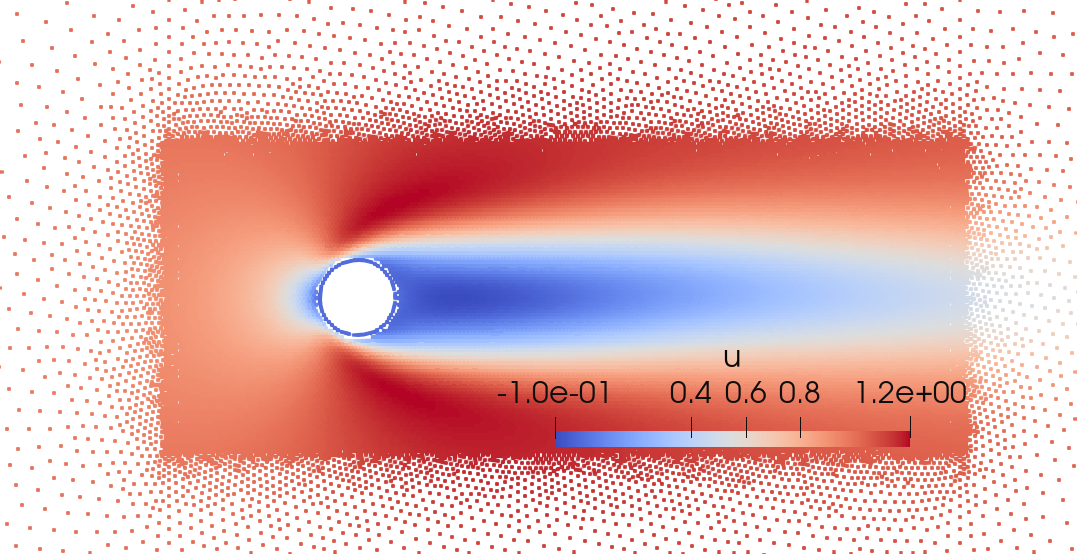}
    }
    \subfloat[]{
    \includegraphics[width=0.45\textwidth]{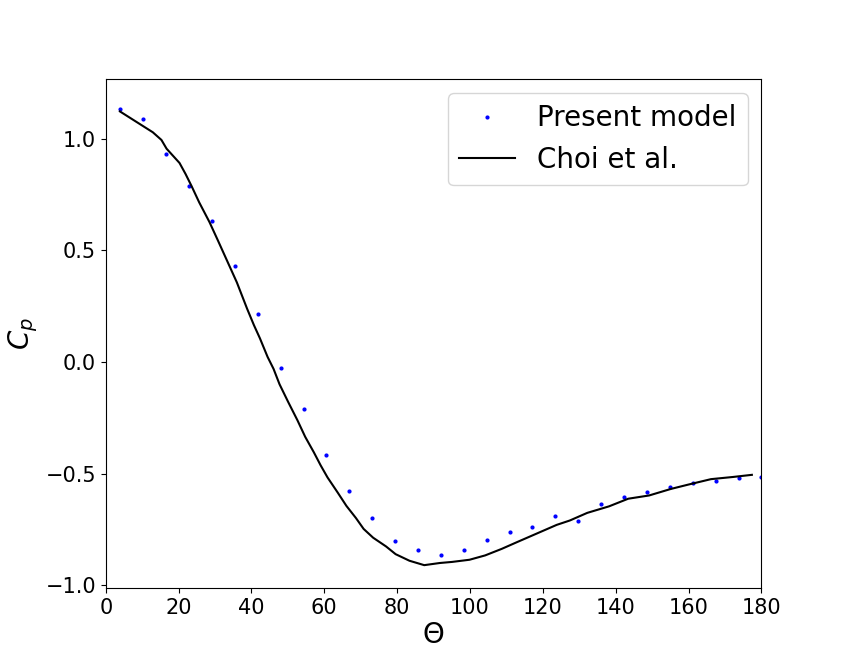}
    }
    \caption{Case 2: Flow past a cylinder; Results for $Re$ = 40; (a) $u$ velocity contours (b) Comparison of $C_p$ with literature }
    \label{fig:C2_solution}
\end{figure}

\subsection{Heat equation on an irregular domain: a convergence study}
In this test case \cite{papac2013level}, we assess the order of accuracy of the method in an irregularly shaped domain by comparing the numerical solution with the analytical solution of the heat equation
\begin{equation}
    \frac{\partial T}{\partial t} = \Delta T\text{, }
\end{equation}
with the Robin boundary condition 
\begin{equation}
    \nabla T \cdot \hat{\mathbf{n}} + T = f.
\end{equation}
The analytical solution is given by
\begin{equation}
    T(x,y,t) = e^{-2t} \cos x \cos y\text{. }
    \label{eq:T1_ana}
\end{equation}
For the numerical simulation, the points in the  domain are generated from a set of uniformly spaced points with the irregular boundary embedded inside the set of points. The embedded boundary is characterized in polar coordinates as 
\begin{equation}
    r = 0.4\cos(8\theta) + \pi, \, \theta \in [0,2\pi]
\end{equation}
The initial condition for the simulation is obtained by setting $t=0$ in Eq.~\ref{eq:T1_ana}. The simulation is performed till $t=1$ unit. At $t=1$, the error in the numerical solution is quantified as
\begin{equation}
    \text{E} =  \frac{1}{N}\sum_{i=1}^N e_i = \frac{1}{N}\sum_{i=1}^N |T^\text{analytical}_i - T^\text{numerical}_i |
\end{equation}
We define smoothing length ($h$) at a point $i$ as the average distance of the points in the neighborhood from point $i$. The neighborhood is chosen to be the 20 nearest neighbors for all the test cases. Therefore, for refined point clouds, the smoothing length is smaller than that for coarser point clouds for the same number of neighbours. Fig.~\ref{fig:Err_convergence} shows the plot of the error versus the smoothing length ($h$) for four different point clouds generated by embedding the irregular boundary inside uniform point clouds of varying resolution. The uniform point clouds consist of $100 \times 100$, $150 \times 150$, $200 \times 200$ and $250 \times 250$ points discretizing a square domain of 8 units length and are identified as $PC_1$, $PC_2$, $PC_3$ and $PC_4$, respectively. The error drops with decreasing smoothing length (i.e. for higher resolution of points) and the order of convergence lies between 1 and 2, as seen. Fig.~\ref{fig:C1_error} shows the error contour plots for three different resolutions of point cloud, $PC_2$, $PC_3$ and $PC_4$. It is seen from the contour legend that the error decreases consistently with increase in resolution. 
\subsection{Flow past a circular cylinder}

This test case verifies a single-phase flow with an embedded surface. We consider the flow past a circular cylinder at a Reynolds number, $Re$ = 40. An illustration of the domain and the boundary conditions are shown in Fig.~\ref{fig:c2_domain}. The domain spans 30$D$ in the stream-wise direction and 20$D$ in the transverse direction, $D$ being the cylinder diameter. The cylinder is placed at 10$D$  from the inflow boundary. 
An initial set of points is obtained from a mesh with a high resolution at the location of the cylinder and its wake. Fig.~\ref{fig:C2_pointcloud}(a) shows this initial point cloud with the cylinder as an embedded surface.  Using the procedure outlined in Sec.~\ref{sec:IB}, a conformal point cloud is generated. Fig.~\ref{fig:C2_pointcloud}(b) and (c) show the entire conformal point cloud and a zoomed-in version of it close to the cylinder, respectively. The points that lie inside the cylinder and just outside the cylinder ($\xi < \xi_{\text{min}}$) are discarded and the surface points are appended to the rest of the points outside ($\xi > \xi_{\text{min}}$), resulting in a body-conforming point cloud. The condition of no-slip is enforced directly at the points that populate the surface of the cylinder. \\
For $Re$ = 40, there is no vortex shedding as is well known and the solution reaches a steady state. Fig.~\ref{fig:C2_solution}(a) shows the $u$-velocity contours of the flow. The wake behind the cylinder is symmetric about the horizontal plane passing through the centre of the cylinder. Fig.~\ref{fig:C2_solution}(b) shows the variation of coefficient of pressure  ($C_p$) along the surface of the cylinder, paramatrized by the angle made with the negative $x$-axis ($\Theta$). It is seen from Fig. \ref{fig:C2_solution}(b) that the present model matches closely with the results of Choi et al.\cite{choi2007immersed}. We reiterate that in the present method, the surface points are directly involved in the discretization and therefore, extracting surface quantities ($C_p$, for example) is quite straightforward, in contrast to traditional immersed-boundary methods where certain interpolation techniques would need to be employed owing to the non-conformal nature of the mesh. \\

\begin{figure}[htb!]
    \centering
    \includegraphics[width=0.6\textwidth]{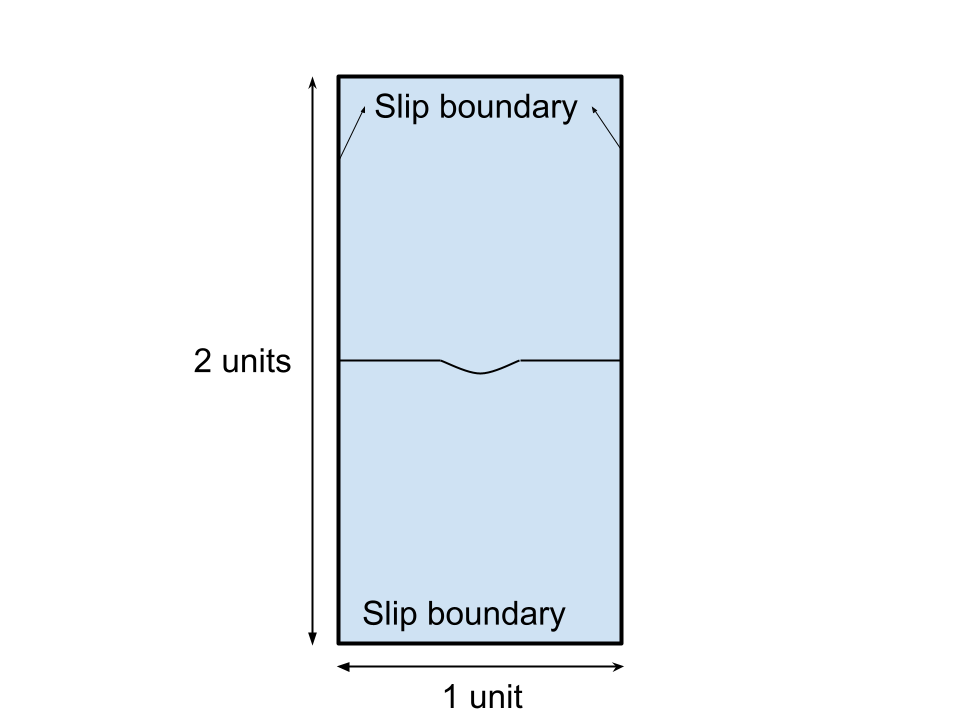}
    \caption{Case 3: Domain and boundary conditions for Rayleigh-Taylor instability.}
    \label{fig:c3_domain}
\end{figure}
\begin{figure}[htb!]
    \centering
    \subfloat[{ $t$=0.4s\raggedright}]{
	    \includegraphics[height=0.5\textwidth,trim={1em 0 0 0}]{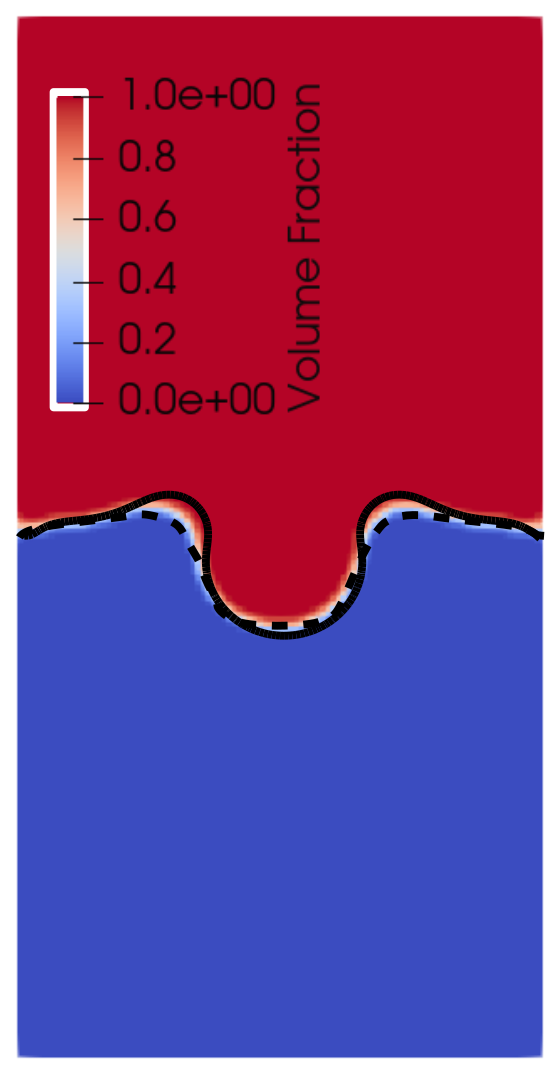}
    }
    \subfloat[{$t$=0.8s\raggedright}]{
    \includegraphics[height=0.5\textwidth,trim={1em 0 0 0}]{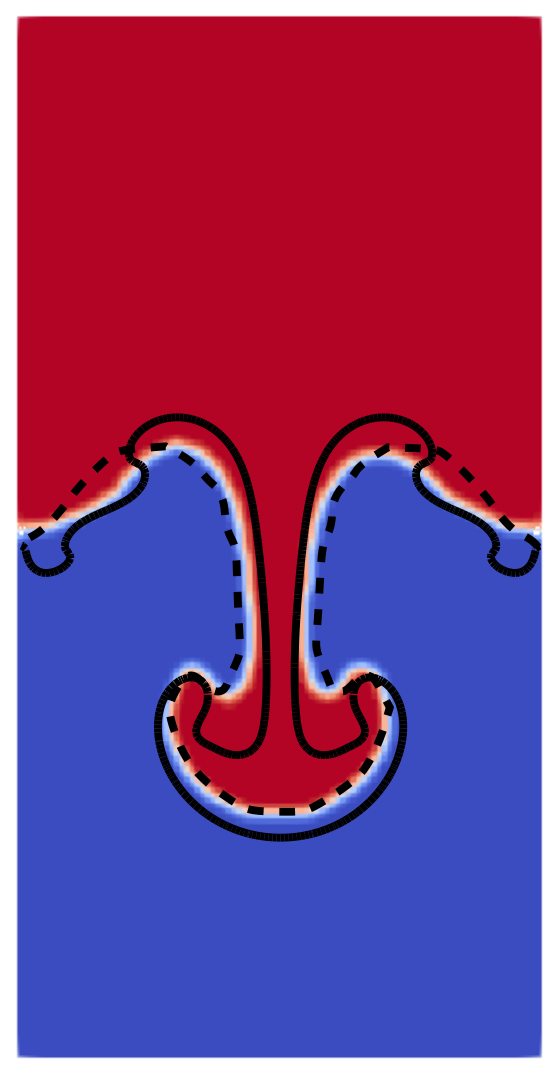}
    }
    \subfloat[{$t$=1.2s\raggedright}]{
    \includegraphics[height=0.5
    \textwidth,trim={1em 0 0 0}]{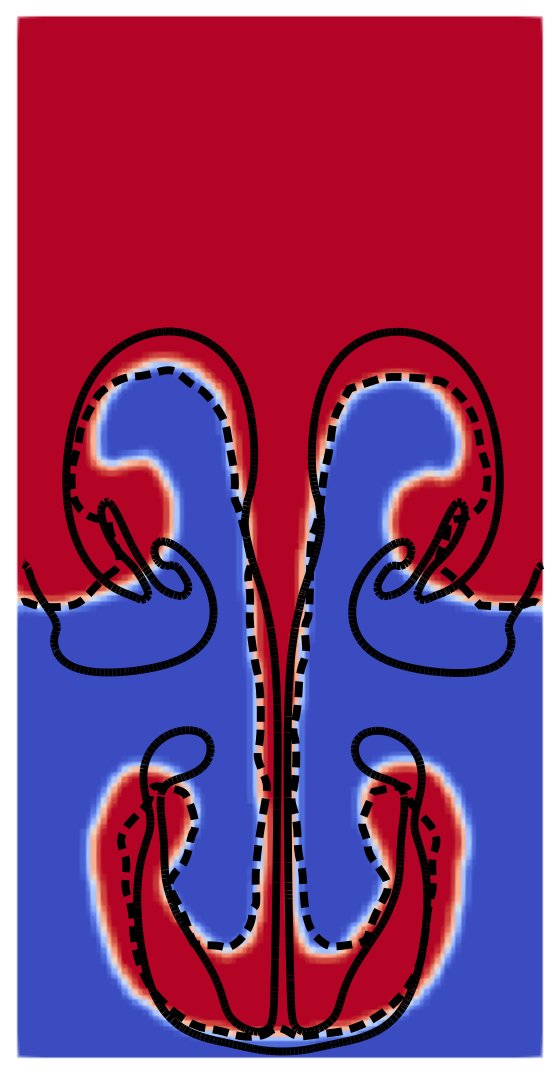}
    }
    \caption{Case 3: Raylor Taylor Instability. The colormap represents the volume fraction of the heavier fluid. The dotted lines indicate the interface position predicted by \cite{liu2005hybrid} and the solid line is the corresponding Finite Volume result using the Gerris \cite{popinet2003gerris} flow solver.}
    \label{fig:C2_RTI}
\end{figure}
\begin{figure}[htb!]
    \centering
    \subfloat[]{
    \includegraphics[width=0.35\textwidth,trim={2cm 0 2cm 0}]{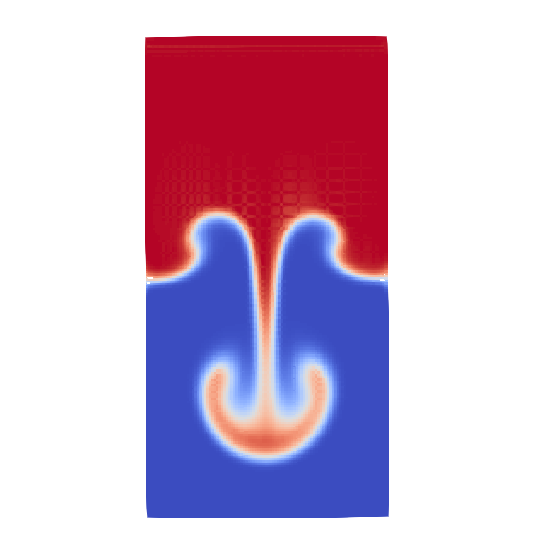}
    }
    \subfloat[]{
    \includegraphics[width=0.35\textwidth,trim={2cm 0 2cm 0}]{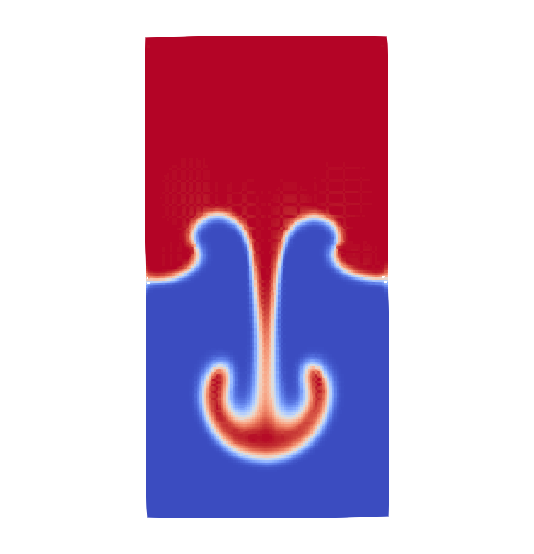}
    }
    \subfloat[]{
    \includegraphics[width=0.35
    \textwidth,trim={2cm 0 2cm 0}]{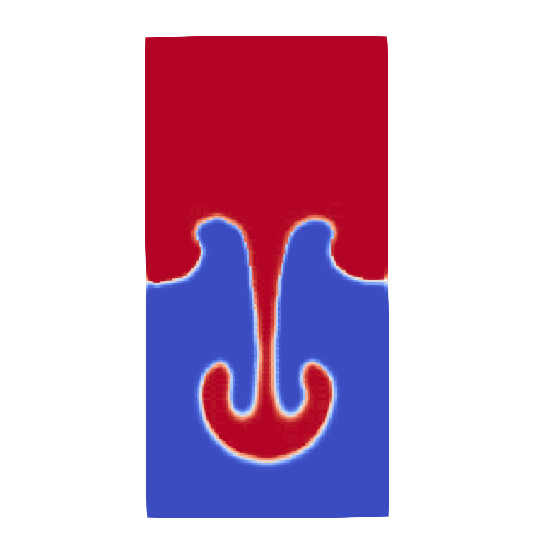}
    }
    \caption{Case 3: Effect of interface sharpening on the solution at $t=1$s; (a) No sharpening (b) sharpening at intervals of $\Delta = 10\%$ (c) sharpening at intervals of $\Delta = 5\%$.}
    \label{fig:C2_RTI_sharpening}
\end{figure}
\subsection{Rayleigh-Taylor Instability}
In this test case, we simulate the Rayleigh-Taylor instability in which a heavier fluid is present on top of a lighter fluid with gravity acting downwards. Since this configuration is inherently unstable, a small perturbation causes the heavier fluid to flow downwards, displacing the lighter fluid. With time, the interface assumes a mushroom-like appearance. \\
The parameters for the simulation are taken with Liu et al. \cite{liu2005hybrid} as the reference. Jeong et al. \cite{jeong2013numerical} and Duan et al.\cite{duan2017stable} also present this test case for purposes of validation. The domain and the boundary conditions are shown in Fig.~\ref{fig:c3_domain} along with a small downward perturbation in the interface. We consider a density ratio, $\rho_h/\rho_l$=3, where $\rho_h$ and $\rho_l$ denote the densities of the heavier and lighter fluids respectively. The kinematic viscosity of both fluids is $\nu_h = \nu_l = 0.01$ $\text{m}^2/\text{s}$ and the surface tension between the fluids is neglected. The same test case is also simulated using the Gerris flow solver \cite{popinet2003gerris}, which uses the finite volume method and the volume of fluid method for interface advection. For the Gerris simulation, we chose a cell resolution of $1/2^7$ m. At $t=0$s, the interface is given a perturbation as shown below 
\begin{equation}
    y =\left\{\begin{aligned}  
       &1.0, && \text{ when } x<0.25 \text{ and } x>0.75 \\ 
       &1 - \delta \sin[2\pi(x-0.25)], && \text{ otherwise}. 
       \end{aligned}
       \right.
\end{equation}
Here $\delta$ is assigned a value of $0.06$.  
Fig.~\ref{fig:C2_RTI}(a)-(c) show the evolution of the interface at different instants of time. At $t=0.4$s, the initial perturbation grows as seen in Fig.~\ref{fig:C2_RTI}(a). At $t=0.8$s, the mushroom-like shape starts to develop as more of the heavier fluid flows downwards (Fig.~\ref{fig:C2_RTI}(b)). At $t=1.2$s, the mushroom-like shape is  prominent as seen in Fig.~\ref{fig:C2_RTI}(c). Both our result and that of \cite{liu2005hybrid} compare well with the Gerris simulation until $t=0.4$s. Owing to greater resolution, the Gerris simulation shows features with greater curvature compared to either of the meshless results. At $t=1.2$s, both the meshless methods predict the mushroom's length reasonably well. However, the volume of the heavier fluid is visibly under-predicted by \cite{liu2005hybrid}.  \\
 Now, we present the effect of interface sharpening introduced in Sec. \ref{sec:sharp} at different frequencies. A measure to evaluate the change in volume fraction ($\Delta$), due to advection, over the entire domain is defined as
\begin{equation}
    \Delta = \frac{1}{N}\sum_i^N |\alpha_i - \alpha_i^{(-)}|.
\end{equation}
Here, the summation is applied over all the points of the domain, $\alpha_i$ denotes the volume fraction at point $i$ and $\alpha_i^{(-)}$ denotes the volume fraction at the last instance of interface sharpening, at the same point. Clearly, $\Delta$ takes a value in the range $[0,1.0]$.
Figs.~\ref{fig:C2_RTI_sharpening}(a)-(c) show the interface at the same instance for three different frequencies of interface sharpening that correspond to $\Delta = 1.0, 0.1 \text{ and } 0.05$ respectively. Higher values of $\Delta$ would imply the sharpening is performed less frequently. It can be seen that the interface becomes quite diffuse when the sharpening is not performed, as seen in Fig.~\ref{fig:C2_RTI_sharpening}(a). The sharpness of the interface improves when the frequency of sharpening is increased by changing $\Delta$ from 0.1 to 0.05, as seen in Fig.~\ref{fig:C2_RTI_sharpening}(b) and (c). It is important to note that reducing $\Delta$ to very low values may lead to a situation where the sharpening algorithm interferes with the natural evolution of the interface according to the advection equation. For this test case and the other two-phase simulations presented in this paper, $\Delta = 0.05$ has been used. 
\subsection{Two-phase Dam Break}
\begin{figure}[htb!]
    \centering
    \includegraphics[width=0.6\textwidth]{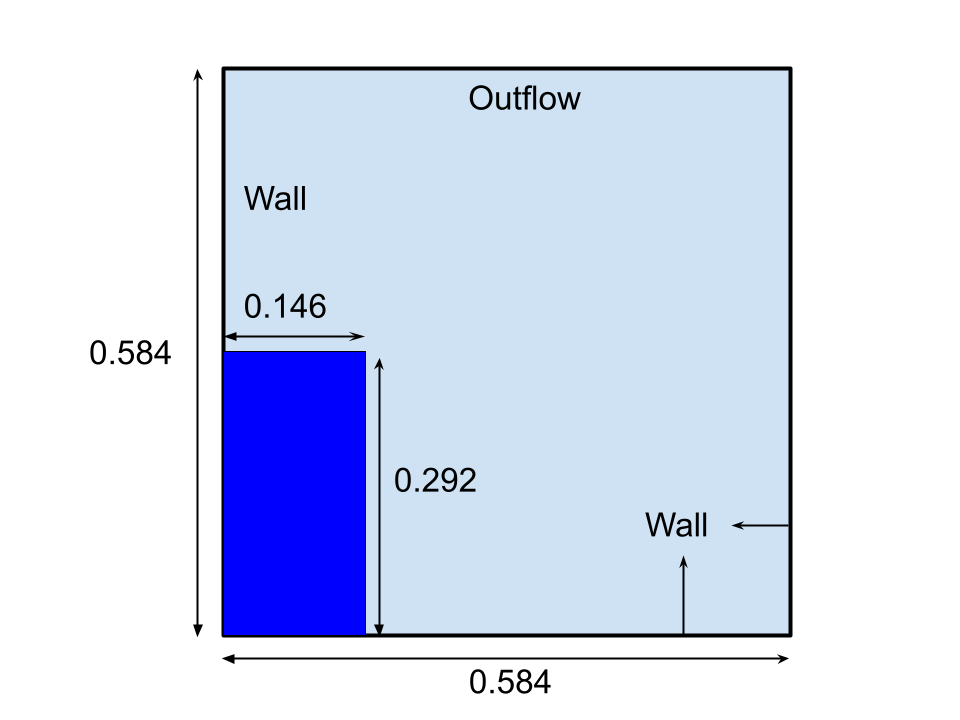}
    \caption{Case 4: Domain and boundary conditions for 2D dam-break problem.}
    \label{fig:c4_domain}
\end{figure}
\begin{figure}[htb!]
    \centering
    \subfloat[t=0.1s]{
    \includegraphics[clip,trim={12cm 1.5cm 12cm 0},width=0.245\textwidth]{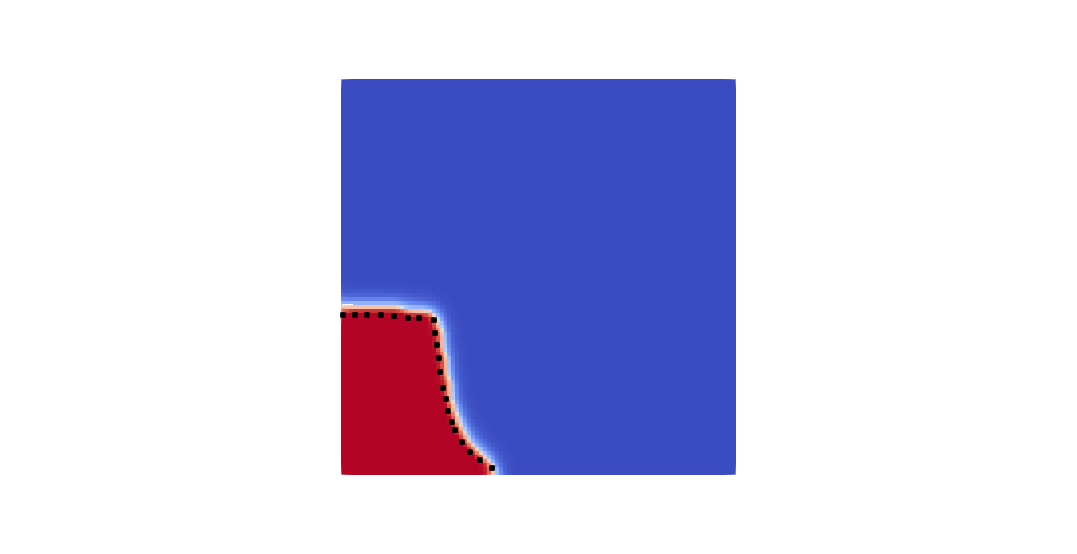}}
    \subfloat[t=0.2s]{
    \includegraphics[clip,trim={12cm 1.5cm 12cm 0},width=0.245\textwidth]{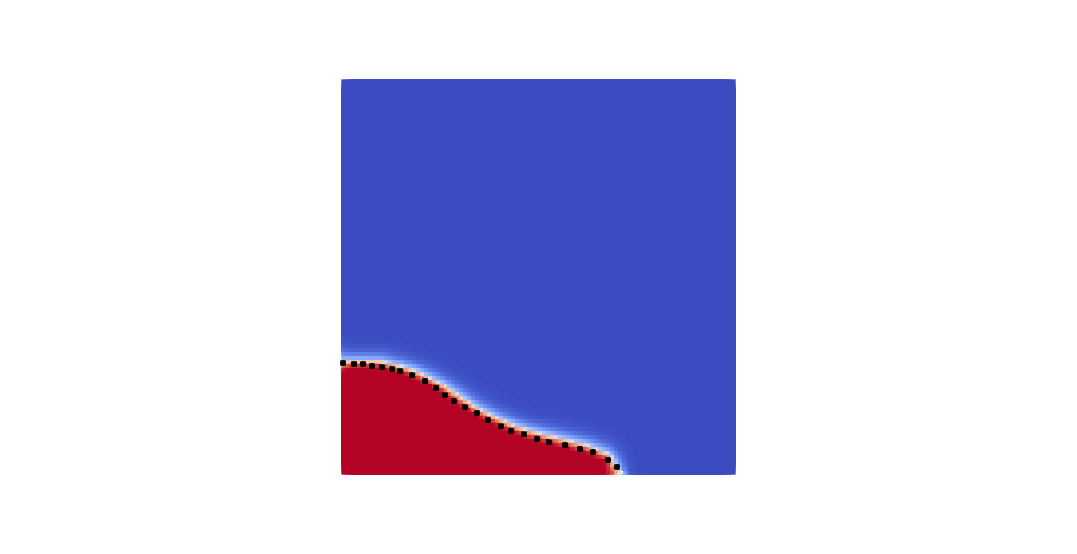}}
    \subfloat[t=0.3s]{
    \includegraphics[clip,trim={12cm 1.5cm 12cm 0},width=0.245\textwidth]{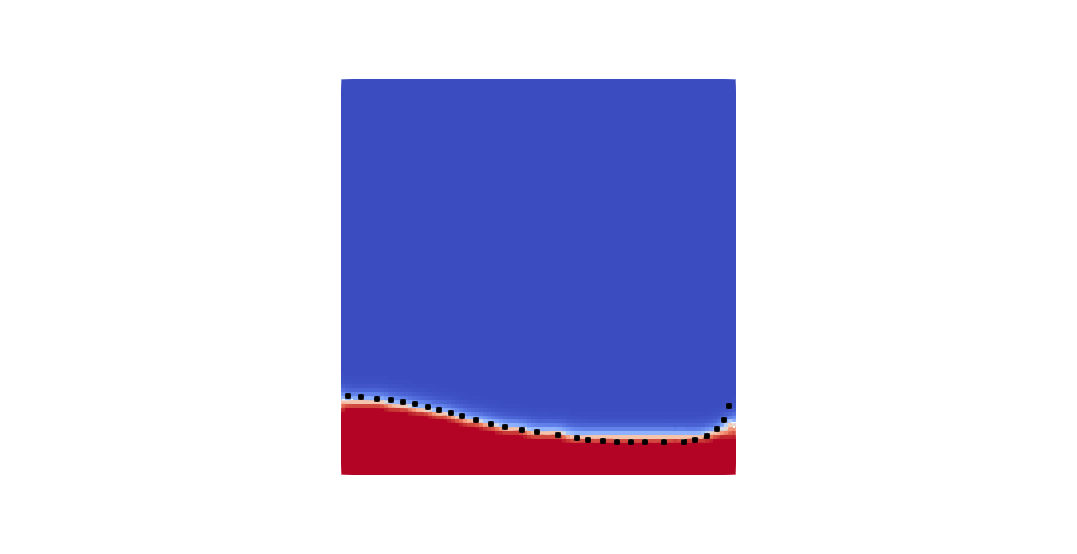}}
    \subfloat[t=0.4s]{
    \includegraphics[clip,trim={12cm 1.5cm 12cm 0},width=0.245\textwidth]{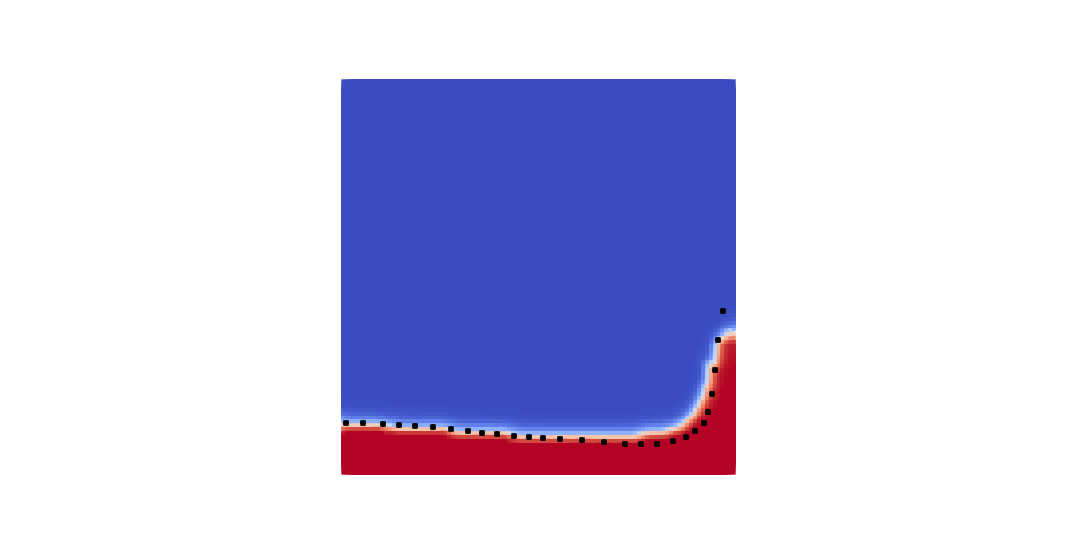}}
    \caption{Case 4: 2D dam-break - Comparison of interface position with Ubbink et. al \cite{ubbink1997numerical} and \cite{jiang2023numerical}. The red contour represents water and blue represents air. The black dots represent the interface obtained from literature.}
    \label{fig:dambreak_comparisons}
\end{figure}
We consider a 2D dam break problem, as illustrated in Fig.~\ref{fig:c4_domain}. The water column (denoted by the dark blue shade) collapses under gravity and we use the proposed method to capture the interface movement. The test case considers water and air as the two phases. The pressure Poisson equation (Eq. \ref{eq:pressurepoisson}) when used for large density ratios ($\rho_{\text{water}}/\rho_{\text{air}} \approx$ 1000), results in numerical instabilities. The gradient of density $\nabla \rho$ is high at the interface when the density ratio is high. We substitute for $\rho$ as \cite{kraus2023meshfree}.
\begin{equation}
    \rho = e^\gamma\text{. }
\end{equation}
Consequently,
\begin{equation}
    \nabla \rho = e^\gamma \nabla \gamma = \rho \nabla \gamma 
\end{equation}\text{. }
The pressure Poisson equation, now, becomes
\begin{equation}
    \nabla^2 p^{n+1} = \frac{\rho}{\Delta t} \nabla \cdot 
    \vec{u}^* +  \nabla p^{n+1} \cdot \nabla \gamma 
    \label{eq:pressurepoisson2}
\end{equation}
The above expression is used instead of Eq.~\ref{eq:pressurepoisson} and is relatively more stable at high density ratios. \\
Fig.~\ref{fig:dambreak_comparisons} shows the evolution of the water-air interface. The black dots are interface positions extracted from the previous work that present the same test case \cite{ubbink1997numerical,jiang2023numerical}. We note a close comparison at $t=0.1$s and $t=0.2$s. At $t=0.3$s and $t=0.4$s, the interface position at the right wall, predicted by the present model is slightly lower than that from the simulations of Ubbink \cite{ubbink1997numerical}. A possible cause is the dissipative error in the numerical advection of the volume fraction (given by Eq.~\ref{eq:alpha_advect}), in which we use a first-order upwind method for reconstruction of fluxes at the fictitious interfaces. Higher order reconstructions are not explored as a part of this work and will be taken up in the future work.
\begin{figure}[htb!]
    \centering
    \subfloat[]{
    \includegraphics[width=0.5\textwidth]{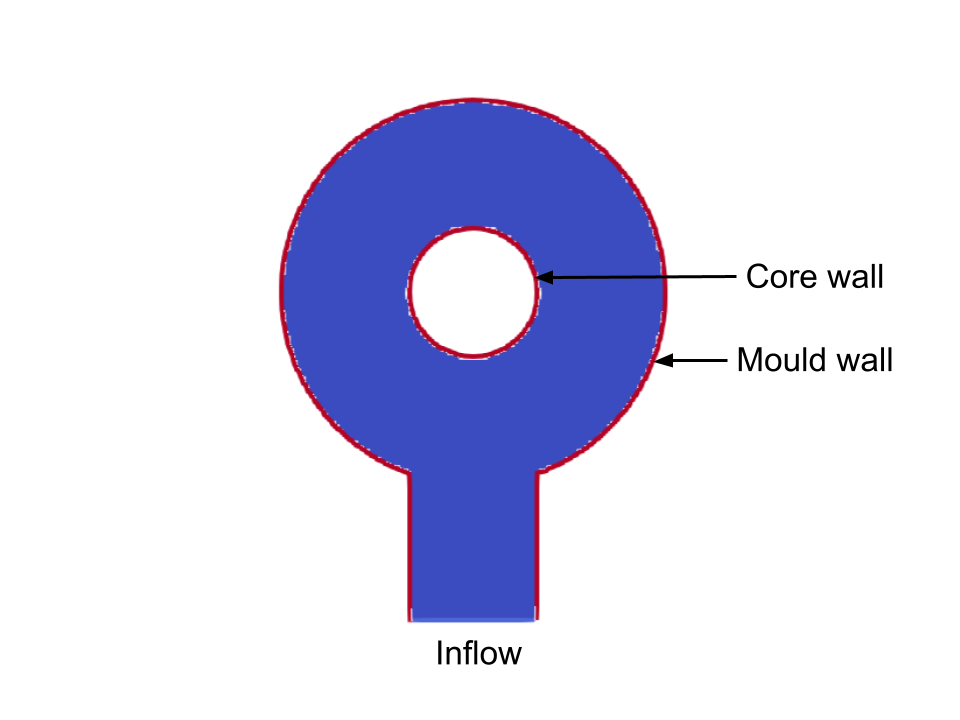}}
    \subfloat[]{
    \includegraphics[width=0.5\textwidth]{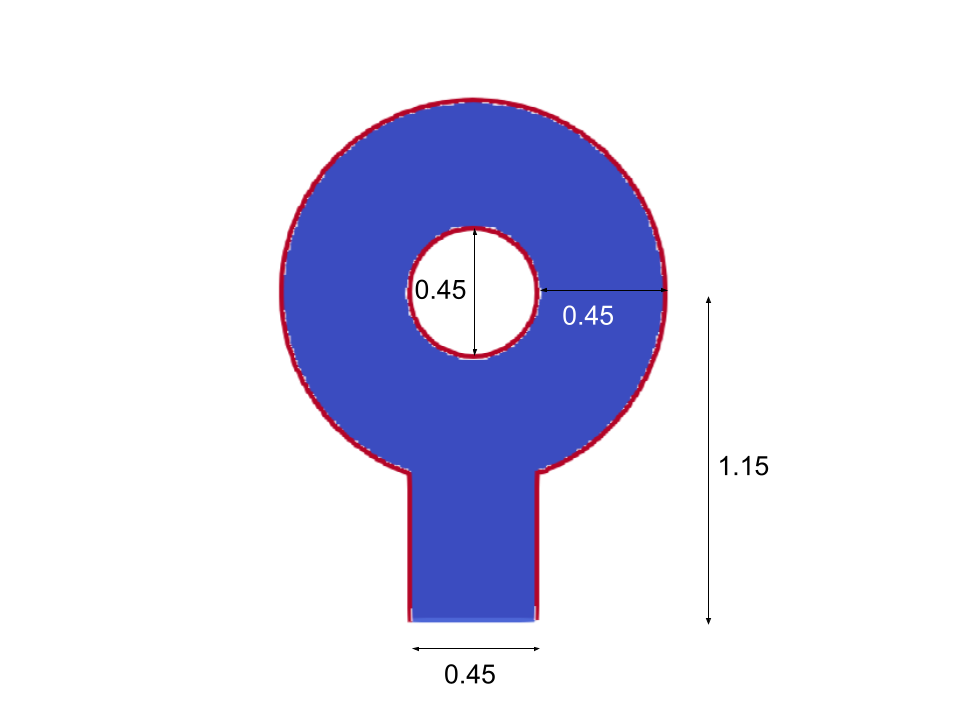}}
    \caption{Case 5: Filling of a mould with circular core (a) Domain and boundary conditions (b) Dimensions of the geometry.}
    \label{fig:c5_domain}
\end{figure}
\subsection{Filling of a mould with core}
\begin{figure}[!htb]
    \centering
    \begin{tabular}{x{0.3\textwidth}x{0.3\textwidth}x{0.3\textwidth}}
	    \textsf{\Large Tracer} & \textsf{\Large U} & \textsf{\Large V}\\
    \end{tabular}\\
    \subfloat[$t$=0.1s]{\includegraphics[trim={13cm 0 5cm 1cm},width=0.3\textwidth]{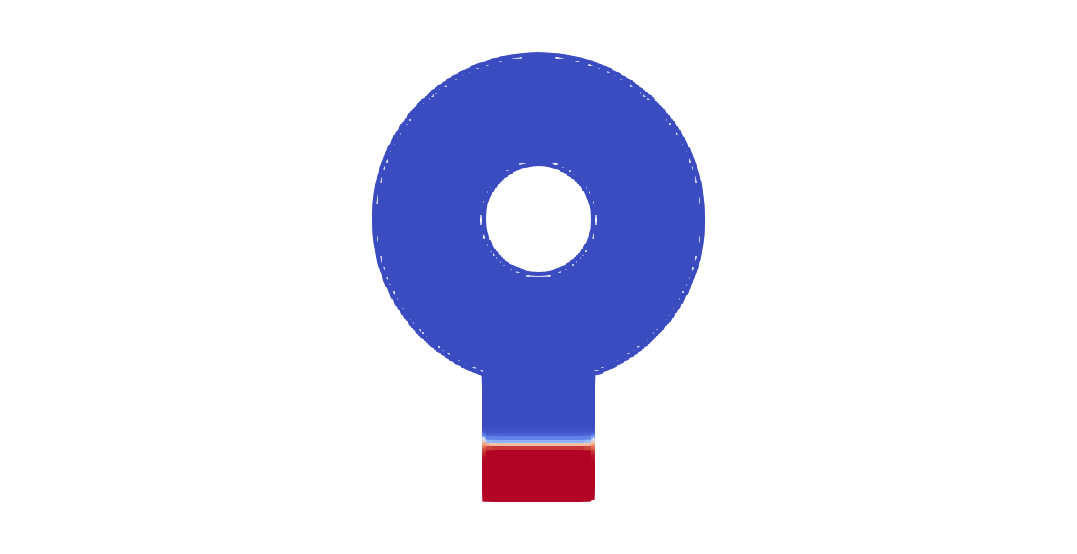}
    \includegraphics[clip,trim={13cm 0 5cm 1cm},width=0.3\textwidth]{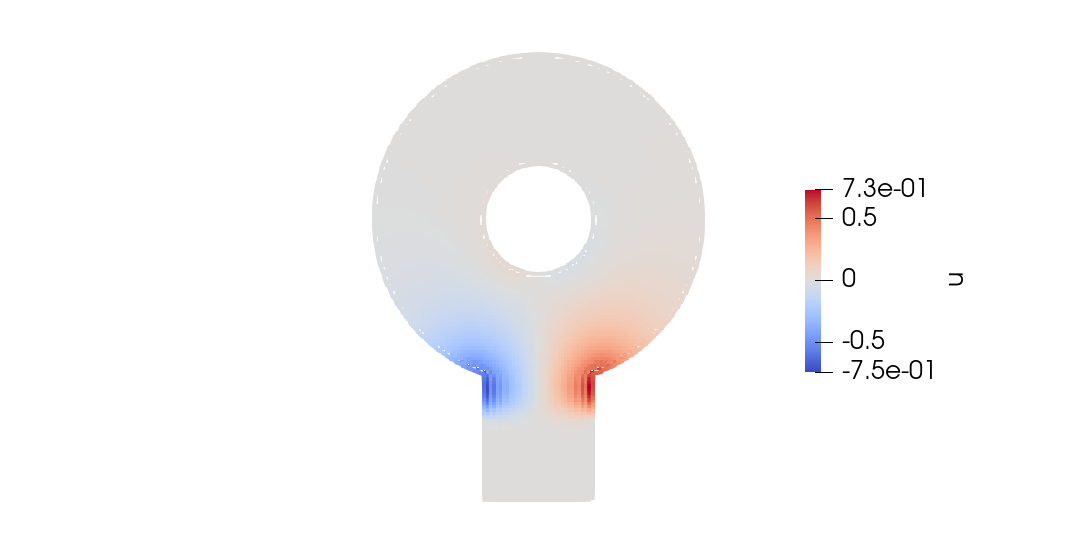}
    \includegraphics[clip,trim={13cm 0 5cm 1cm},width=0.3\textwidth]{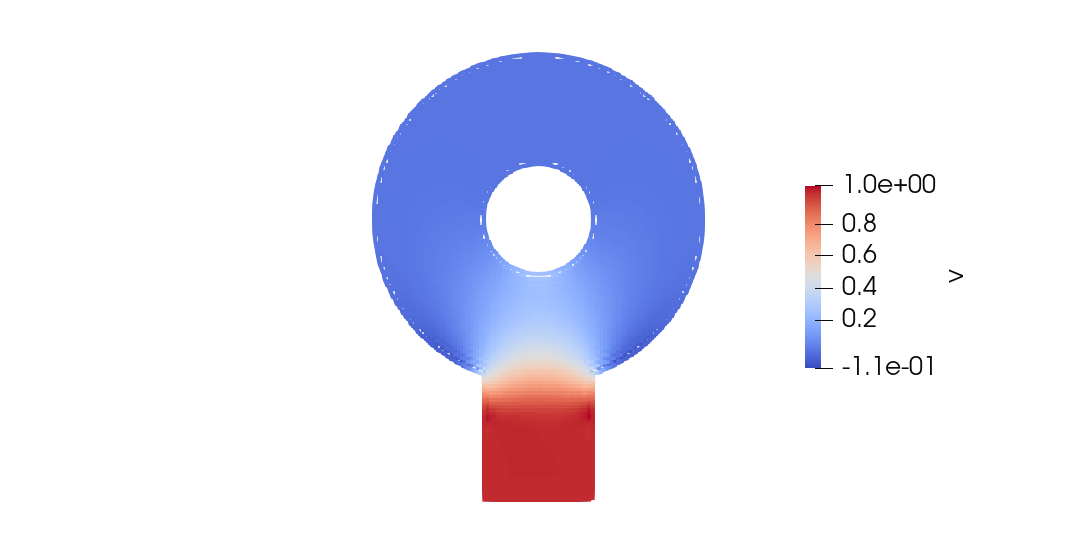}
    } \\
    \subfloat[$t$=0.5s]{\includegraphics[trim={13cm 0 5cm 1cm},width=0.3\textwidth]{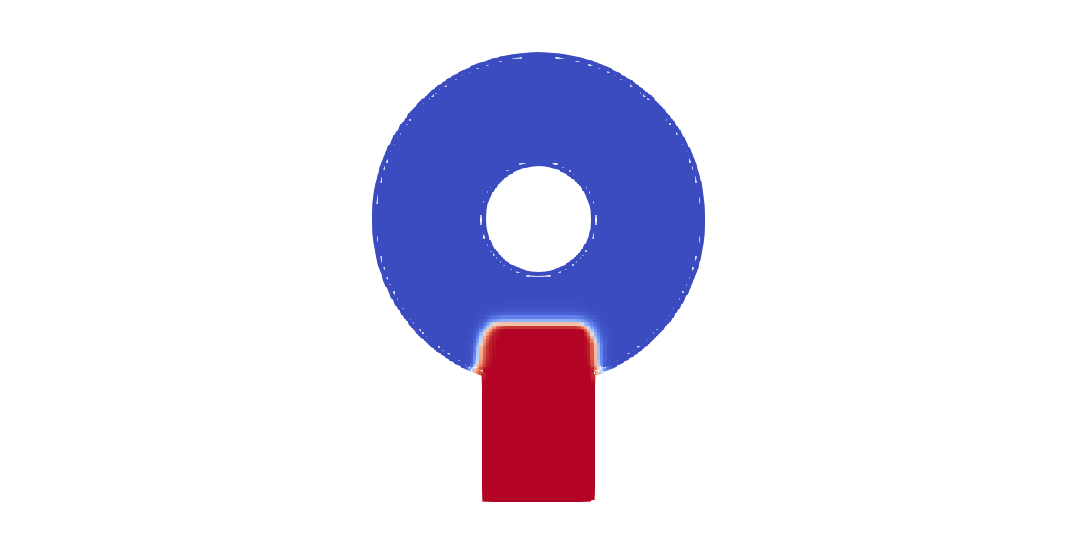}
    \includegraphics[clip,trim={13cm 0 5cm 1cm},width=0.3\textwidth]{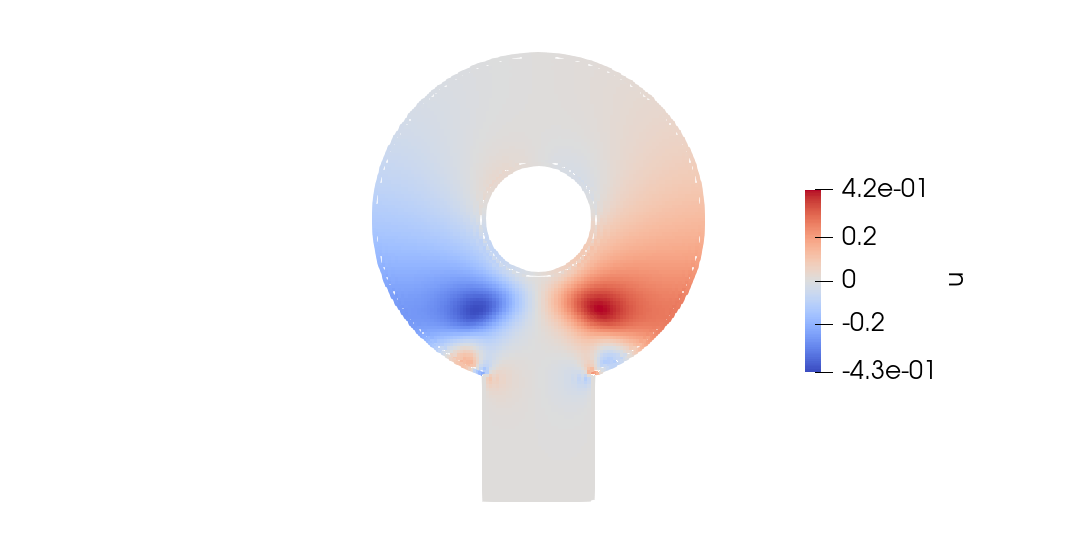}
    \includegraphics[clip,trim={13cm 0 5cm 1cm},width=0.3\textwidth]{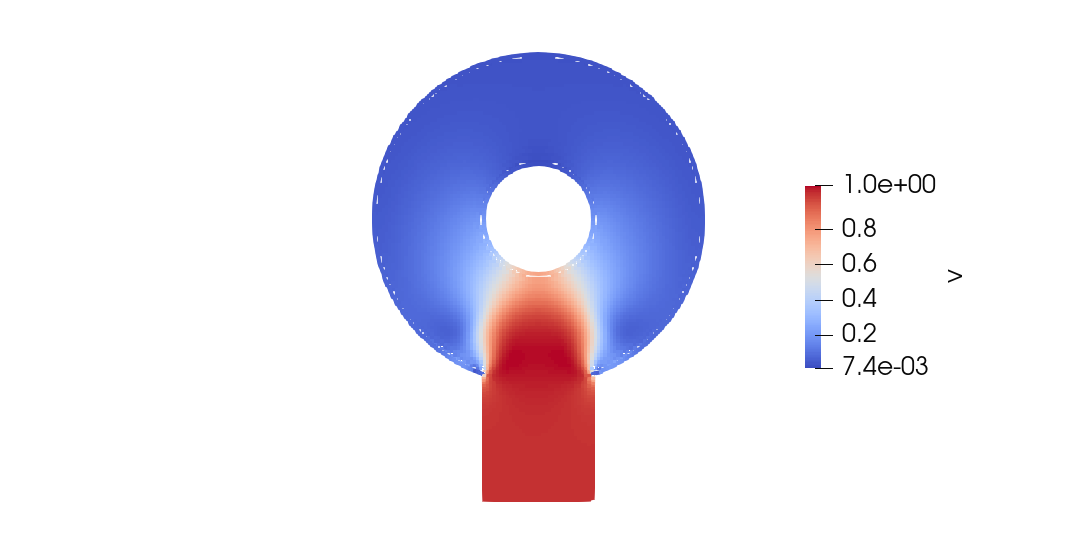}
    }
     \\
    \subfloat[$t$=1s]{\includegraphics[trim={13cm 0 5cm 1cm},width=0.3\textwidth]{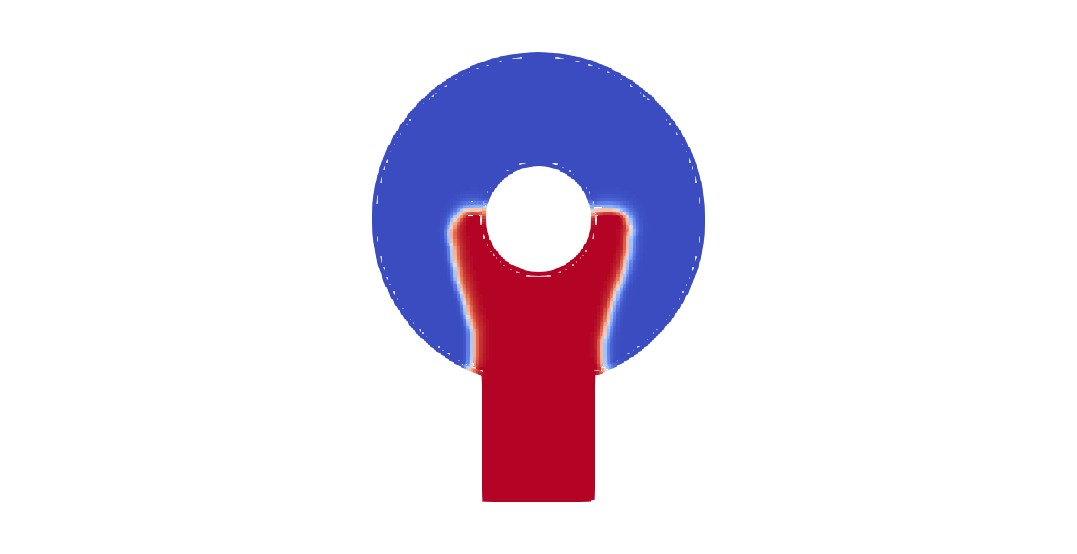}
    \includegraphics[clip,trim={13cm 0 5cm 1cm},width=0.3\textwidth]{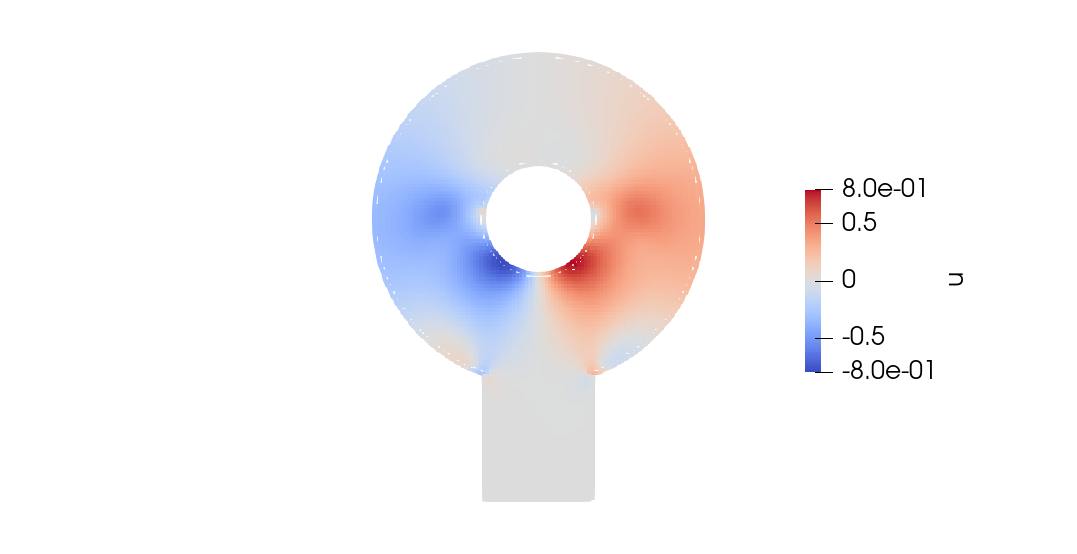}
    \includegraphics[clip,trim={13cm 0 5cm 1cm},width=0.3\textwidth]{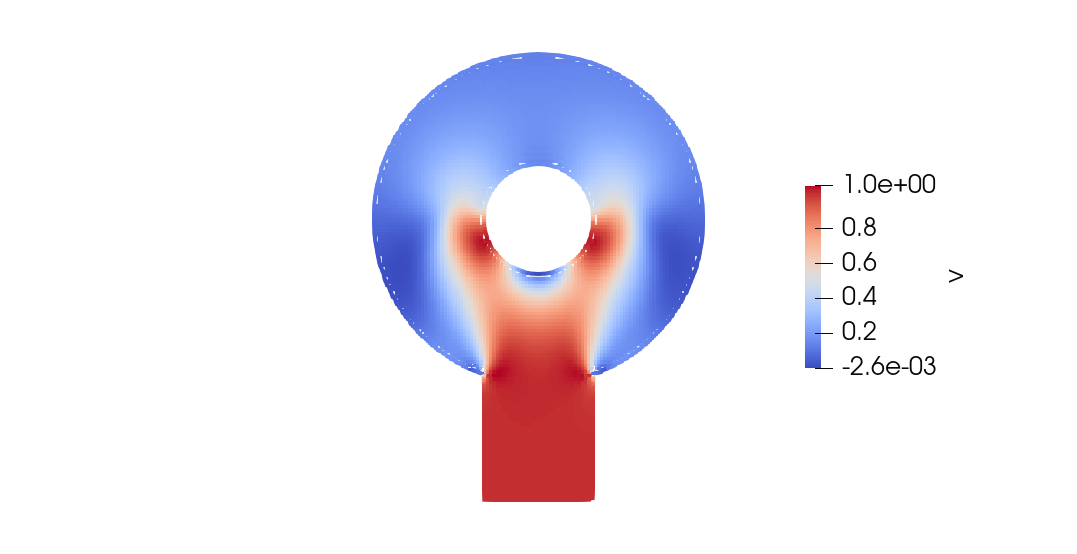}
    }\\
    \caption{Case 5: The filling of liquid in a mould cavity with a circular core at different instances of time. The left column shows the two phases - red denotes liquid and blue denotes gas. The middle column shows the $u$-velocity contours and the right column shows the $v$-velocity contours. Continued.}
    \end{figure}
    \begin{figure}[!htb]
    \ContinuedFloat
    \centering
    \subfloat[$t$=1.5s]{\includegraphics[trim={10cm 0 5cm 0},width=0.3\textwidth]{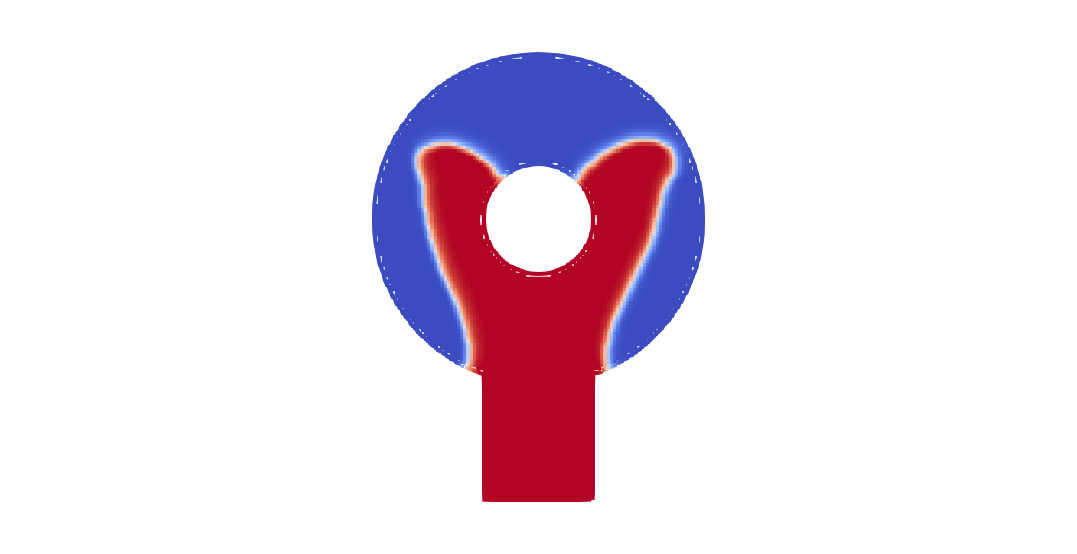}
    \includegraphics[trim={10cm 0 5cm 0},width=0.3\textwidth]{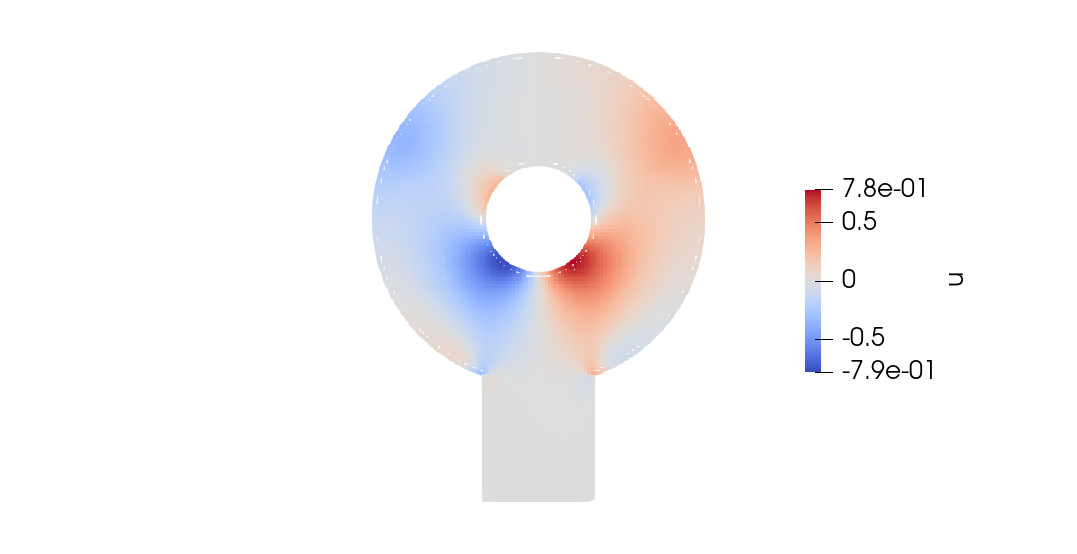}
    \includegraphics[trim={10cm 0 5cm 0},width=0.3\textwidth]{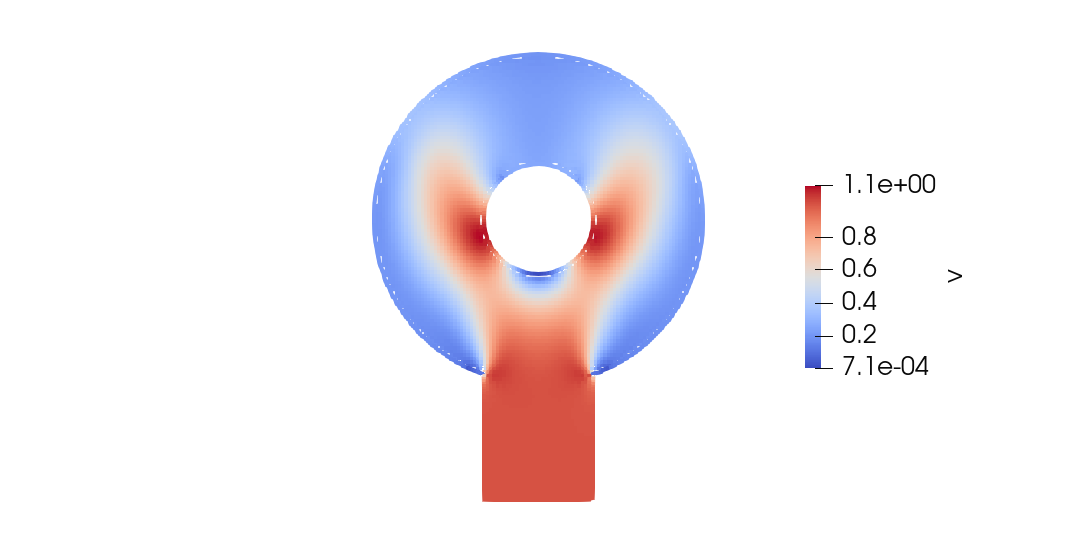}
    }\\
    \subfloat[$t$=2s]{\includegraphics[trim={10cm 0 5cm 0},width=0.3\textwidth]{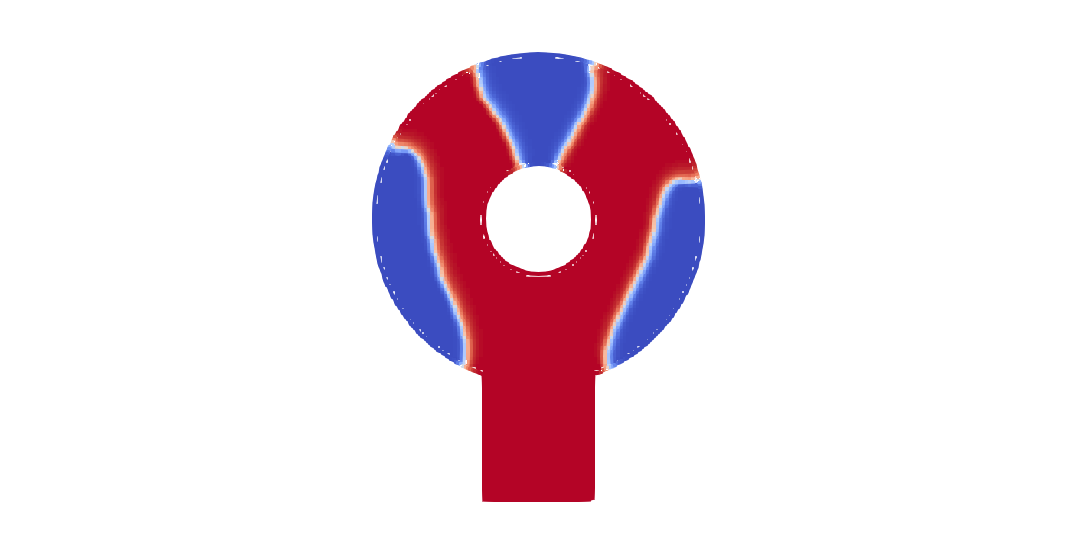}
    \includegraphics[trim={10cm 0 5cm 0},width=0.3\textwidth]{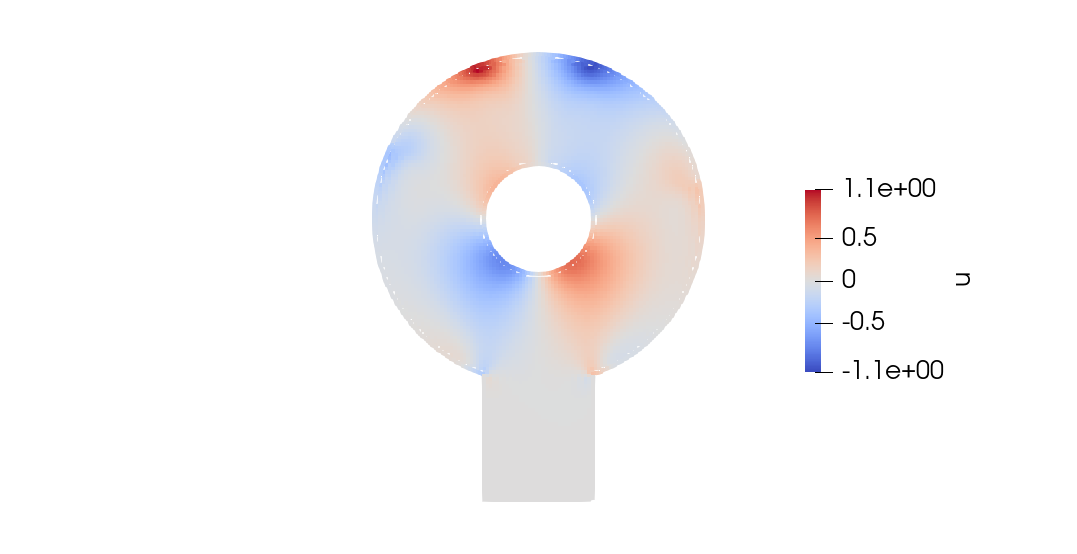}
    \includegraphics[trim={10cm 0 5cm 0},width=0.3\textwidth]{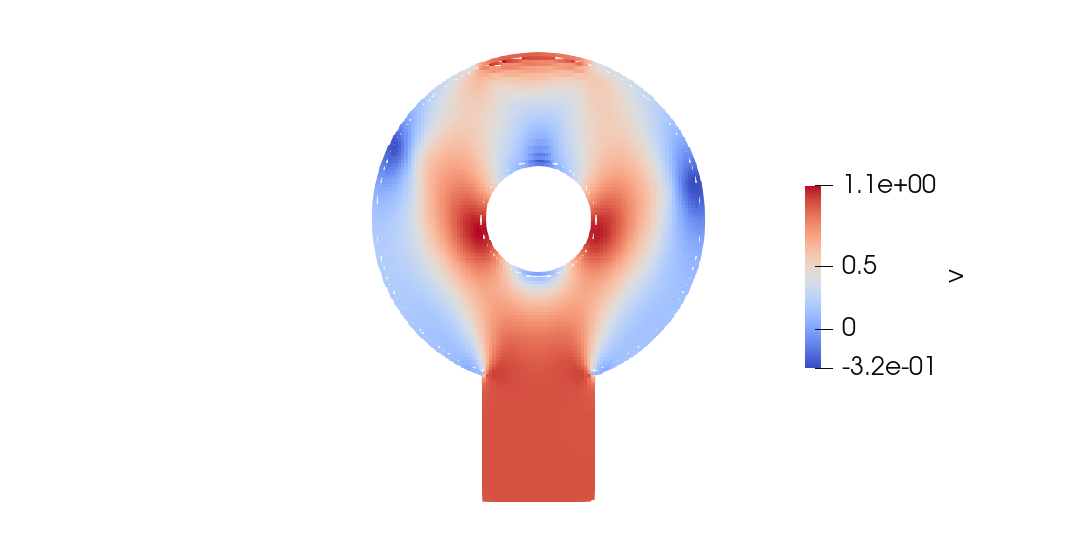}
    }\\
    \subfloat[$t$=2.5s]{\includegraphics[trim={10cm 0 5cm 0},width=0.3\textwidth]{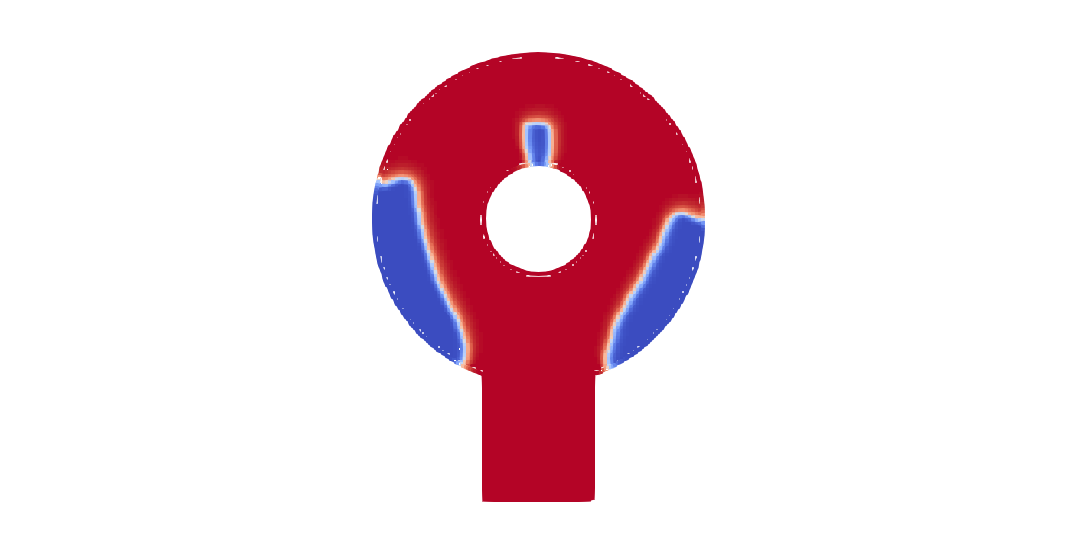}
    \includegraphics[trim={10cm 0 5cm 0},width=0.3\textwidth]{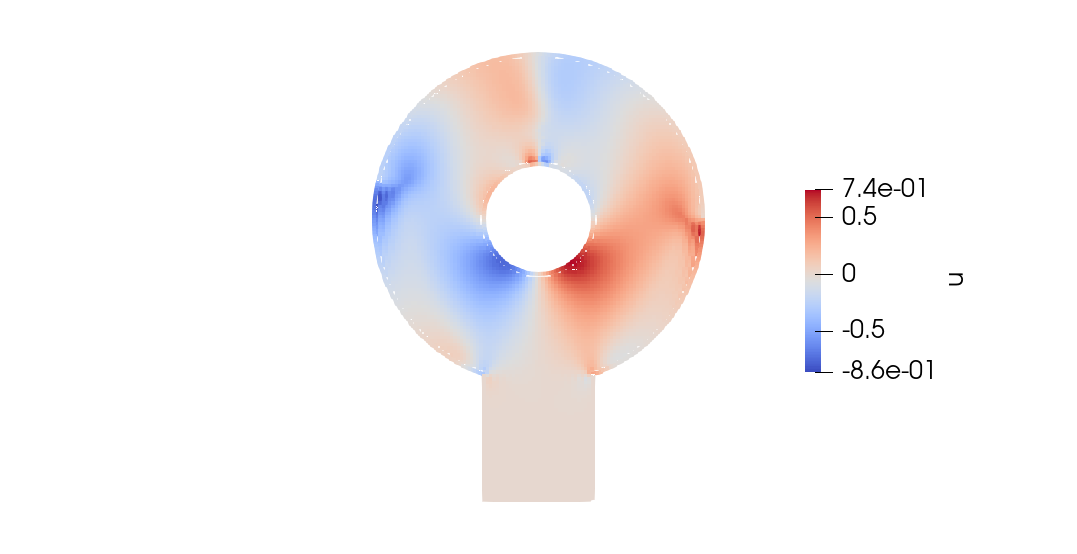}
    \includegraphics[trim={10cm 0 5cm 0},width=0.3\textwidth]{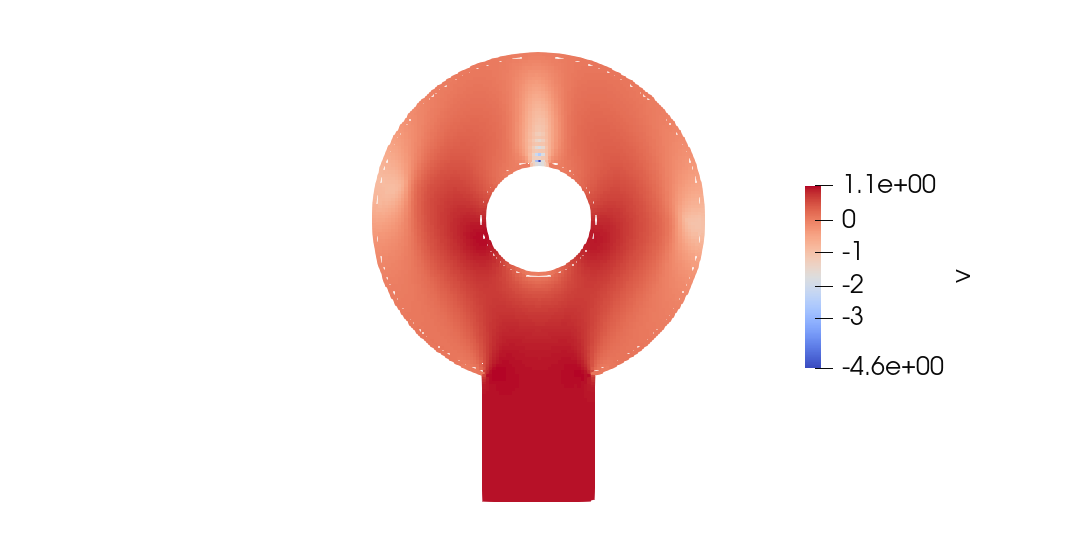}
    }\\
    \caption{Case 5: The filling of liquid in a mould cavity with a circular core at different instances of time. The left column shows the two phases - red denotes liquid and blue denotes gas. The middle column shows the $u$-velocity contours and the right column shows the $v$-velocity contours.}
    \label{fig:C5_results}
\end{figure}
In this test case, we consider a flow problem that the present method is targeted to solve i.e. two-phase flows with embedded geometries. We consider a mould with a circular core as the flow domain, as shown in Fig.~\ref{fig:c5_domain}(a). The dimensions of the geometry are shown in Fig.~\ref{fig:c5_domain}(b) \cite{saucedo2019three}. Starting from a uniform set of points, the conformal point cloud is generated using the method described in Sec.~\ref{sec:IB}. The ratio of the density of liquid to that of the gas ($\rho_l/\rho_g$) of 100 and a Reynolds number of 100 (based on the inflow liquid velocity and unit length), are considered. A time-step $\Delta t =$ 0.0001s is used and the distance between two adjacent points of the uniform point cloud is $\delta x =$ 0.01m. $\delta x$ is also used as the resolution for the population of points on the embedded surfaces. At the mould and core walls, we apply a boundary condition that depends on the volume fraction. When the volume fraction is zero, denoting gas phase, the walls act as outflow such that the gas flows out without any resistance. On the other hand, when the volume fraction is unity, denoting liquid, the fluid encounters the slip-wall condition. \\
Fig.~\ref{fig:C5_results}(a)-(f) show the filling of the mould cavity at different instants of time. The left column of the figure shows the two phases, with red denoting the liquid (the heavy fluid) and blue denoting the gas (the light fluid). The middle and the right columns show the $u$ and $v$ velocity contours. The liquid enters the mould cavity from the bottom (Fig.~\ref{fig:C5_results}(a) and (b)). As this happens, the gas is forced out of the mould cavity as seen in the velocity contours. When the liquid jet impinges on the core, it splits the jet into two as shown in Fig.~\ref{fig:C5_results}(c) and (d). The jets, subsequently, reach the mould wall as seen Fig.~\ref{fig:C5_results}(e). The liquid flows along the wall as it continues to fill the cavity, as seen in Fig.~\ref{fig:C5_results}(f). This test case, motivated by its application to casting industry, demonstrates the feasibility of our method for two phase flows with embedded boundaries. 

\section{Conclusions}\label{sec:conclusions}
This paper presents an Eulerian meshless solver for two-phase flows with embedded geometries. The advantages of both the Eulerian framework and the meshless framework are captured in this solver. Owing to the Eulerian framework, neighbourhood search and the differential operator calculations are not required to be performed at every timestep, as is the case with Lagrangian methods. The meshless aspect of the model retains the advantage that a prior mesh is not necessary for the computation. Additionally, the meshless framework would be advantageous when point cloud adaptations are performed, as planned in  future extensions of the method. \review{ The limitation of using an Eulerian approach is that the interface separating the phases needs to be tracked using a field variable, requiring explicit mass conservation. In contrast, in the Lagrangian approaches the interface evolves naturally as the points move ensuring better mass conservation.}\\
The two-phase model uses a volume fraction to track the phase and the interface movement is captured through the advection of the volume fraction. The solution of the advection equation in the meshfree framework requires the use of a direction flux-based minimization procedure, elaborated in this paper. An interface sharpening approach is proposed as an auxiliary method to retain the sharpness of the interface. A method to generate point clouds that are conformal to arbitrary geometries starting with a non-conformal point cloud, is also proposed. \\
The test cases validate the model for both two-phase flows and flows with embedded geometries. The final test case which involves the filling of a mould with a core, illustrates the use of the model for flows with two phases as well as embedded geometries. \\
\review{Future directions would include enhancing the accuracy of the method, incorporating surface tension forces in the model, and using adaptation of the point cloud in the vicinity of the interface and  the vicinity of embedded geometries.} Two-phase flow through porous cavities and filling of dies in casting manufacturing are potential areas of application of the present model. 
\section*{Acknowledgements}
Anand S Bharadwaj would like to acknowledge the Science and Engineering Research Board (SERB) for funding  through the National Post Doctoral Fellowship Scheme. Prapanch Nair would like to acknowledge the SERB for funding through the Startup Research Grant SRG/2022/000436.
\bibliographystyle{unsrt}
\bibliography{main.bib}
\end{document}